\documentclass[aps,pre,twocolumn,groupedaddress]{revtex4-2}

\usepackage{graphicx}
\usepackage{amsmath}
\usepackage{amssymb}
\usepackage{amsfonts}
\usepackage{hyperref}

\begin{document}


\title{Anomalous Behavior in Systems with Delay}


\author{Tony Albers}
\email[]{tony.albers@physik.tu-chemnitz.de}
\author{David M\"uller-Bender}
\email[]{david.mueller-bender@mailbox.org}
\author{Lukas Hille}
\email[]{hille.lukas@proton.me}
\author{Martin Weigel}
\email[]{martin.weigel@physik.tu-chemnitz.de}
\affiliation{Institute of Physics, Chemnitz University of Technology, 09107 Chemnitz, Germany}


\date{\today}

\begin{abstract}
In a recent letter [Phys. Rev. E \textbf{112}, L042201 (2025)],
we demonstrated the occurrence of anomalous diffusion, weak chaos, and weak ergodicity breaking in a certain class of time-delayed feedback systems with linear instantaneous and nonlinear delayed term.
This anomalous behavior is caused by the trapping of chaotic solutions close to nonhyperbolic fixed-point solutions in function space.
In this paper, we extend our previous study, present more detailed insights, and show surprising new results.
We first investigate the case of a constant delay, where the dynamical behavior during such trapping events can be explained by center manifold theory.
In the large-delay limit, the probability for the trapping events becomes small because of a high effective dimension of the chaotic phases.
As a consequence, anomalous behavior is only observable on very large time scales.
By introducing a periodic modulation of the delay time, trapping events become more probable with increasing modulation amplitude due to a reduction of the effective dimension and the occurrence of laminar chaos.
For large values of the modulation amplitude, the anomalous behavior is more complex and the most relevant aspects can be reproduced by a two-dimensional iterated map and a stochastic model.
\end{abstract}


\maketitle

\section{\label{sec:1}Introduction}

Chaotic diffusion \cite{geisel1982,schell1982,fujisaka1982} is a purely deterministic but erratic type of motion typically occurring in nonlinear dynamical systems exhibiting chaos without any randomness.
While the state variable in such systems often shows normal chaotic diffusion,
the appearance of anomalous diffusion - usually defined by a nonlinear increase of the mean-squared displacement \cite{bouchaud1990,klages2008} - is of special interest.
The mechanisms leading to such anomalous chaotic diffusion have been intensively investigated and are well understood for low-dimensional iterated maps \cite{geisel1984,zumofen1993,bel2006}
and Hamiltonian systems \cite{chirikov1979,zacherl1986,geisel1987,lichtenberg1992,zumofen1994}.
If one considers dynamical systems with finite reaction and transmission times, however, the dimension of the problem becomes infinite
and the dynamics have to be described by delay differential equations \cite{hale1993,diekmann1995}.
Such time-delays naturally occur in the description of dynamical systems in many contexts such as
classical electrodynamics (finite propagation time) \cite{driver1963,lopez2020}, population dynamics (time to sexual maturity) \cite{gopalsamy1992,kuang1993},
and epidemiology (incubation time) \cite{salpeter1998,kajiwara2012}, to name but a few.
Furthermore, they often play a role in different application areas such as nonlinear optics \cite{ikeda1980,erneux2009} and engineering \cite{erneux2009,insperger2011}.
Therefore, delay systems are of rather fundamental importance.
Regarding the occurrence of chaotic diffusion in such infinite-dimensional systems, only a few such cases have been reported to date, most of them dealing with normal diffusion
\cite{wischert1994,schanz2003,sprott2007,lei2011,dao2013_1,dao2013_2,mackey2021,albers2019_1,albers2019_2,albers2022_1,albers2022_2}.
There are only two examples where anomalous chaotic diffusion has been observed in the context of partial differential equations \cite{cisternas2016,cisternas2018}.
The type of anomalous diffusion reported there and which we also find in the present study is subdiffusive.
Subdiffusion has been observed in several systems such as transport of charge carriers in amorphous semiconductors \cite{scher1975},
diffusion of molecules on the plasma membrane \cite{schuetz1997,krapf2015}, and diffusion on fractal structures like percolation clusters \cite{gefen1983,havlin1987}.
It is often described by means of stochastic models such as continuous-time random walks \cite{montroll1965,metzler2000} and fractional Brownian motion \cite{mandelbrot1968}.

In a previous work \cite{albers2025}, some of us reported on the occurrence of anomalous chaotic diffusion in systems with delay for the first time.
Moreover, we found weak chaos \cite{korabel2009,korabel2010}, i.e., the maximal Lyapunov exponent is equal to zero,
and weak ergodicity breaking \cite{bouchaud1992,lubelski2008,he2008} meaning that ensemble and time averages do not coincide, where the latter become random variables.
In the present work, we extend and generalize the results reported in Ref.~\cite{albers2025}.
For instance, we demonstrate that in the case of a modulated delay with large modulation amplitude, the occurrence of the above mentioned anomalous behavior is more complex than one would naively expect.
We are then able to reduce the complex behavior to such an extent that it can be described by a two-dimensional iterated map and a simple stochastic model.

The rest of this paper is organized as follows.
In Sec.~\ref{sec:2}, we introduce the delay system under consideration in its most general form.
We then start our investigation with the special case of a constant delay described in Sec.~\ref{sec:3}, where a more analytical approach is possible using the theory of the center manifold.
We then continue with an investigation of the more general case of a time-varying delay in Sec.~\ref{sec:4}.
A special emphasis lies on the investigation of the maximal Lyapunov exponent in Sec.~\ref{sec:5}.
In Sec.~\ref{sec:6}, we focus on the case of a high modulation amplitude of the delay and describe the dynamical behavior with an iterated map in Sec.~\ref{sec:7} and a stochastic model in Sec.~\ref{sec:8}.
A brief summary and conclusion are given in Sec.~\ref{sec:9}.

\section{\label{sec:2}Delay System}

As in the previous studies \cite{albers2022_1,albers2022_2,albers2025}, the class of time-delayed feedback systems under consideration is determined by the scalar delay differential equation (DDE) of the form
\begin{equation}
\label{eq:DDE}
\frac{1}{\Theta}\dot{x}(t)=-x(t)+f\boldsymbol{(}x[R(t)]\boldsymbol{)}
\end{equation}
with a linear instantaneous and a nonlinear delayed term.
The parameter $\Theta>0$ specifies the overall time scale and $R(t)$ is the retarded argument defined by
\begin{equation}
\label{eq:R_t}
R(t)=t-\tau(t).
\end{equation}
For the time-varying delay
\begin{equation}
\label{eq:tau_t}
\tau(t)=\tau_0-\frac{A}{2\pi}\sin(2\pi t),
\end{equation}
which we consider here, a mean delay $\tau_0>0$ is sinusoidally modulated with modulation amplitude $A\geq0$.
This choice contains the special case of a constant delay $\tau(t)=\tau_0$ for $A=0$.
Obviously, the time-varying delay $\tau(t)$ has to be positive for every time $t$ meaning that $A<2\pi\tau_0$.
Because we are interested in the influence of the strength of the periodic delay modulation, in the following we set the mean delay equal to unity, $\tau_0=1$,
and change the modulation amplitude in the range $0\leq A<1$ such that the retarded argument $R(t)$ is monotonically increasing.
The time scale transformation $t'=\Theta t$ together with the definition $y(t'):=x(t'/\Theta)$ changes Eq.~(\ref{eq:DDE}) into $\dot{y}(t')=-y(t')+f\boldsymbol{(}y[t'-\Theta\tau(t'/\Theta)]\boldsymbol{)}$
demonstrating that the case $\Theta\gg1$ is equivalent to the large delay limit.
The latter is typically of particular interest in the literature \cite{ikeda1982,chow1983,mallet-paret1986,ikeda1987,mensour1998,wolfrum2006,adhikari2008,wolfrum2010,lichtner2011,giacomelli2012,marino2014,amil2015}
and is also in the focus of our investigation.
A specific choice of the nonlinear function $f(x)$ completes the definition of the considered time-delayed system.
Particular examples studied in the literature \cite{lakshmanan2011} include the Mackey-Glass equation, $f(x)=\mu x/(1+x^{10})$, \cite{mackey1977} and the Ikeda equation, $f(x)=\mu\sin(x)$, \cite{ikeda1980,ikeda1982}.
In previous works \cite{albers2022_1,albers2022_2}, we used the so-called climbing-sine nonlinearity $f(x)=x-\mu\sin(2\pi x)$ \cite{geisel1982} leading to normal chaotic diffusion.
For this choice, the resulting DDE possesses only unstable hyperbolic fixed-point solutions in function space.
In the present work, we want to develop a deeper understanding of the influence of additional unstable nonhyperbolic fixed-point solutions on the dynamics and especially on the chaotic diffusion.
To do so, we use as a natural extension of the climbing-sine nonlinearity the so-called double-sine nonlinearity \cite{albers2025}
\begin{equation}
\label{eq:f_x}
f_{\mu}(x)=x-\mu[\sin(2\pi x)+0.5\sin(4\pi x)],
\end{equation}
which shows reflection $f_{\mu}(-x)=-f_{\mu}(x)$ and discrete translational symmetry $f_{\mu}(x\pm1)=f_{\mu}(x)\pm1$.
While the former excludes the occurrence of an additional drift, the latter allows the division of the real axis into unit cells.
The resulting time-delayed system, Eqs.~(\ref{eq:DDE})-(\ref{eq:f_x}), possesses unstable hyperbolic fixed-point solutions at the integer values of $x$
and especially also unstable nonhyperbolic fixed-point solutions at the half integers.
To be more precise, we first perform a linear stability analysis of these fixed-point solutions in the case of a constant delay, $\tau(t)=\tau_0=1$:
We start with the nonhyperbolic fixed points $x_{\text{nh}}^*(t)=x_{\text{nh}}^*=1/2+n\in\mathbb{Z}$ and consider a slightly perturbed solution of the DDE $\tilde{x}(t)=x_{\text{nh}}^*+\delta x(t)$.
Inserted into Eq.~(\ref{eq:DDE}), linearized around $x_{\text{nh}}^*$, and using $f_{\mu}'(x_{\text{nh}}^*)=1$, this leads to the linear DDE
\begin{equation}
\label{eq:lin_DDE}
\frac{1}{\Theta}\delta\dot{x}(t)=-\delta x(t)+\delta x(t-1).
\end{equation}
The corresponding characteristic equation is given by $\lambda/\Theta=-1+e^{-\lambda}$.
One solution of this equation, i.e., one stability exponent, is equal to zero, while all the others are complex with negative real parts \cite{gushchin1999}.
A corresponding calculation for the hyperbolic fixed points $x_{\text{h}}^*(t)=x_{\text{h}}^*=n\in\mathbb{Z}$ leads to the characteristic equation $\lambda/\Theta=-1+(1-4\pi\mu)e^{-\lambda}$.
In the large delay limit ($\Theta\gg1$), a finite number of the complex solutions have positive real part while all the others have negative real part.
For the case of time-varying delay, we numerically find that the largest stability exponent of the non-hyperbolic fixed points is also equal to zero
and for the hyperbolic fixed points, it also has positive real part.

We want to conclude this section with some general remarks on the DDE of Eq.~(\ref{eq:DDE}).
Starting from an initial function $x_0(t)$ defined on the time interval $(t_{-1}=R(t_0),t_0]$, the DDE in Eq.~(\ref{eq:DDE}) can be solved stepwise by a procedure called the method of steps \cite{bellman1965}:
By introducing solution segments $x_n(t)$ defined on state intervals $(t_{n-1}=R(t_n),t_n]$, the solution on the $(n+1)$st state interval can be calculated from the solution on the $n$th state interval
\cite{ikeda1987,mueller2018,mueller2019},
\begin{equation}
\label{eq:mos}
x_{n+1}(t)=x_n(t_n)e^{-\Theta(t-t_n)}+\int_{t_n}^t\Theta e^{-\Theta(t-t')}f\boldsymbol{(}x_n[R(t')]\boldsymbol{)}\,\text{d}t'.
\end{equation}
Because the current state of the delay system is given by the $n$th solution segment $x_n(t)$, i.e., a function on the $n$th state interval, such delay systems are infinite-dimensional.
It is these solution segments that may diffuse in function space but because of their finite variation, one can reduce the consideration to the diffusive motion of one component of the solution segments, e.g.,
the value in the middle of each solution segment.
The solution $x(t)$ itself can thereby be interpreted as an unbounded phase variable.
For the following numerical simulations, we solved the DDE in Eq.~(\ref{eq:DDE}) by using the two-stage Lobatto IIIC method with linear interpolation and step size $\Delta t=0.001$
as described in Ref.~\cite{bellen2003}.

\section{\label{sec:3}Constant Delay: Center manifold}

We start our investigation with the simpler case of a constant delay ($A=0$) before later on proceeding to the time-varying delay from Eq.~(\ref{eq:R_t}) and Eq.~(\ref{eq:tau_t}) in Sec.~\ref{sec:4}.

\begin{figure}
\includegraphics[width=\linewidth]{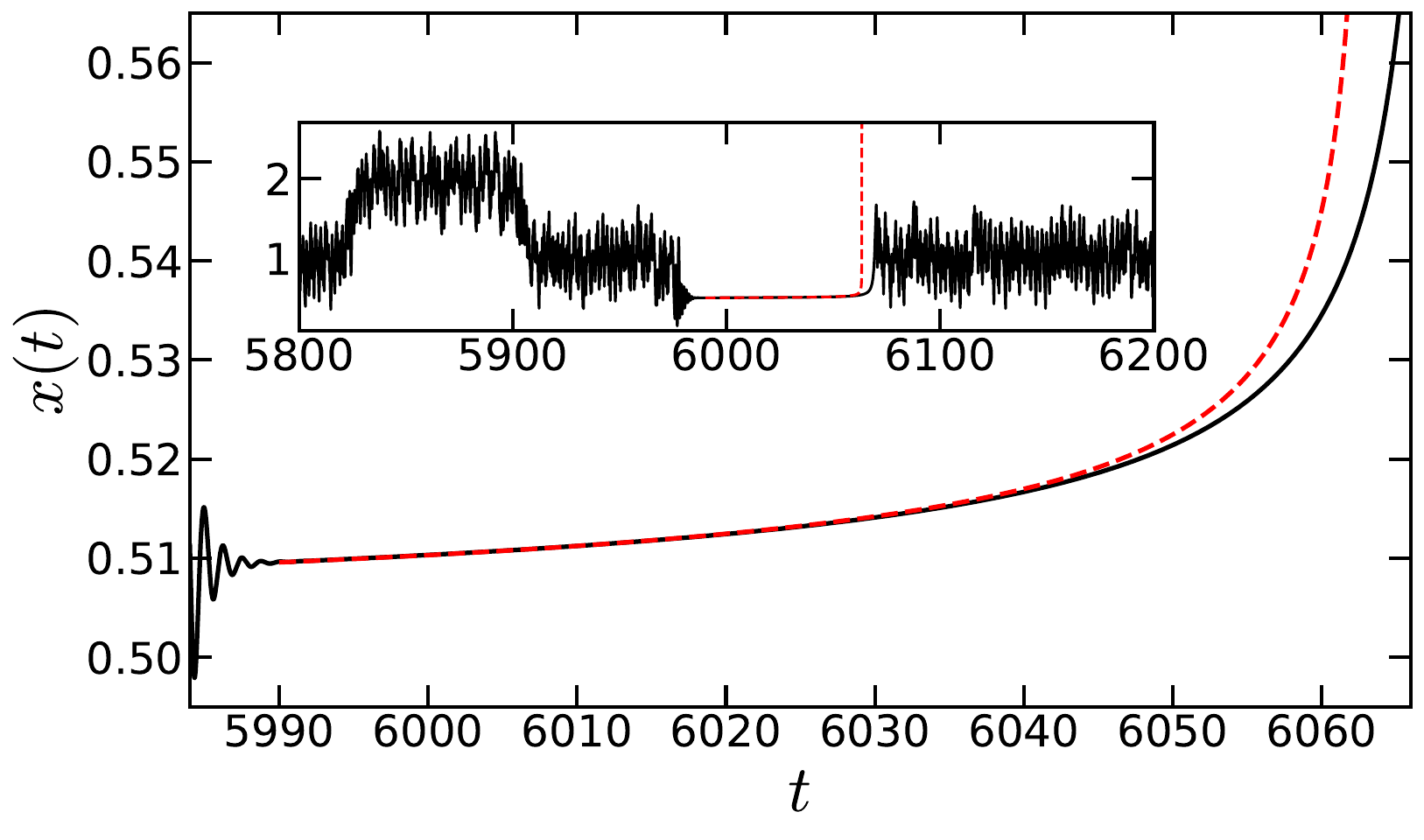}
\caption{\label{fig:1}
Trapping of a chaotic solution of the DDE of Eqs.~(\ref{eq:DDE})-(\ref{eq:f_x}) for a constant delay ($\Theta=2$, $\tau_0=1$, $A=0$, $\mu=0.9$)
close to the unstable nonhyperbolic fixed-point solution $x_{\text{nh}}^*(t)=0.5$ leads to a laminar phase.
The black curve is the numerical solution of the DDE shown on a small time scale in the main figure and on a larger time scale in the inset.
The red curve is the approximation obtained by center manifold theory.}
\end{figure}

\begin{figure}
\includegraphics[width=\linewidth]{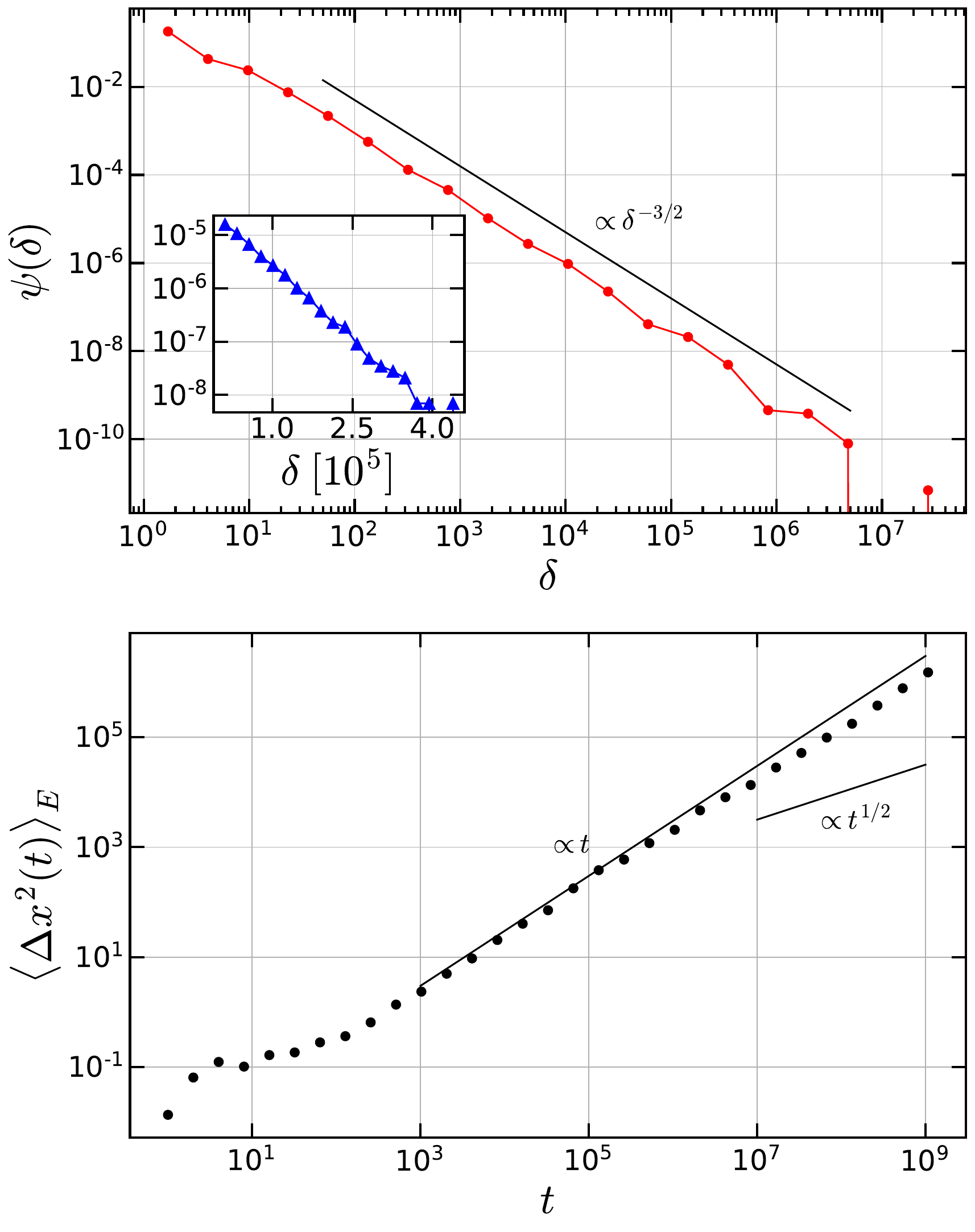}
\caption{\label{fig:2}
The upper main figure and inset show that the distributions of the durations of laminar phases and turbulent phases are heavy-tailed and exponential, respectively.
The MSD in the lower part of the figure increases almost linearly except for the last part, where a slight deviation from the linear behavior indicates a transition to anomalous subdiffusion.
The simulation parameters are the same as in Fig.~\ref{fig:1}.}
\end{figure}

Fig.~(\ref{fig:1}) shows a single numerically determined chaotic solution of the DDE on a shorter and, in the inset, on a longer time scale.
On the longer scale, one recognizes that turbulent phases, named in accordance to the term ``optical turbulence'' in \cite{ikeda1980}, are interrupted by laminar phases of almost constant value
\footnote{We define the solution in one state interval to be laminar if the Euclidean distance of the corresponding solution segment from the nearest marginally unstable fixed-point solution is smaller than $0.05$.}.
The latter are caused by the trapping of chaotic solutions close to the unstable nonhyperbolic fixed-point solutions.
Such trapping of chaotic solutions close to regular solutions for quite a while is a dynamical phenomenon that is especially known from Hamiltonian systems \cite{karney1983,ichikawa1987,zaslavsky2002}.
The approaching of chaotic solutions to the unstable nonhyperbolic fixed-point solutions is due to their stability exponents having negative real part.
The subsequent slow removal is due to the largest stability exponent being equal to zero and can be understood in more detail with center manifold theory that will be discussed later in this section.
We studied the observed dynamical behavior statistically and found that the durations of the turbulent phases follow an exponential distribution,
whereas the distribution of the durations of the laminar phases is heavy-tailed with exponent consistent with $-3/2$ meaning that the mean value of this distribution diverges.
This is illustrated in the upper panel of Fig.~\ref{fig:2}.
During the turbulent phases, the diffusion is normal \cite{albers2022_1}, but these phases are interrupted by the laminar phases, which can be interpreted as waiting times.
From a random-walk point of view, the situation is similar to the subdiffusive continuous-time random walk (CTRW).
In this model, instantaneous jumps randomly drawn from a distribution with finite second moment are interrupted by heavy-tailed distributed waiting times with diverging mean value.
As a consequence, the overall diffusion process is subdiffusive.
However, for our delay system, this subdiffusion can hardly be seen in the mean-squared displacement (MSD), which is shown in the lower part of Fig.~\ref{fig:2}.
The reason is that subdiffusion occurs only asymptotically after a preasymptotic linear increase of the MSD
because of the very large mean value of the distribution of durations of turbulent phases (here roughly $\langle\Delta_{t}\rangle\approx50000$).
With the help of the stochastic model to be discussed in Sec.~\ref{sec:8}, the transition time between normal and subdiffusion in the MSD can be estimated as $t^*\approx(\langle\Delta_t\rangle/\sqrt{\pi})^2$.
For the case shown in Fig.~\ref{fig:2}, this transition time is about $t^*\approx8\cdot10^8$.
In general, it is known \cite{farmer1982} that the dimension of our delay system increases linearly with $\Theta$.
Correspondingly, the probability that a chaotic solution is trapped by an unstable nonhyperbolic fixed-point solution decreases exponentially with $\Theta$
because for a trapping event every point in one solution segment has to come close to the fixed-point solution.
As a consequence, the mean value of the distribution of chaotic phases increases exponentially with $\Theta$.
Therefore, the transition time between normal diffusion and subdiffusion also increases exponentially with $\Theta$, meaning that for larger values of $\Theta$ it is practically impossible to observe the subdiffusion.

As already mentioned, the underlying explanation for the anomalous behavior are the laminar phases whose durations follow a distribution with a heavy tail.
For the constant-delay case, this power-law behavior can be derived analytically using the center manifold theory \cite{hale1993,diekmann1995}.
In the following, we present this derivation.

We consider time-delay systems with the structure
\begin{equation}
\label{eq:sys}
\dot{x}(t)=Lx_t+f(x_t),
\end{equation}
where the solution $x(t)$ is written in Hale-Krasovskii notation $x_t(t'):=x(t+t')$ with $t\in\mathbb{R}$ and $t'\in[-\tau,0]$.
The right-hand side consists of the linear part of the DDE given by $Lx_t$ and the nonlinear part given by $f(x_t)$,
which fulfills $f(0)=0$ and $(Df)(0)=0$, where $D$ is the derivative in the underlying function space $C$.
For example, our DDE from Eq.~(\ref{eq:DDE}) with constant delay $\tau(t)=\tau_0=1$ and the nonlinearity from Eq.~(\ref{eq:f_x}) can be approximated close to the nonhyperbolic fixed points $x_{\text{nh}}^*$ by
\begin{equation}
\label{eq:sys_our}
\dot{x}(t)=\Theta\{-x(t)+x(t-1)+4\pi^3\mu[x(t-1)]^3\},
\end{equation}
where $x(t)$ is the distance to $x_{\text{nh}}^*$.
Eq.~(\ref{eq:sys_our}) can be converted into Eq.~(\ref{eq:sys}) by defining $Lx_t=\Theta[-x_t(0)+x_t(-1)]$ and $f(x_t)=\Theta4\pi^3\mu[x_t(-1)]^3$.

The following theory on the dynamics of time-delay systems on the center manifold can be found in large detail in \cite[Chapter 10.2]{hale1993} and in a different form also in \cite[Chapter IX]{diekmann1995}.
Let us consider the linearization of Eq.~(\ref{eq:sys}) around the equilibrium $x_t=0$, which is given by
\begin{equation}
\label{eq:sys_lin}
\dot{x}(t)=Lx_t.
\end{equation}
Neglecting small and other pathological solutions, the solution space $C$ can be decomposed as $C=U\oplus N\oplus S$,
where the spaces $U$, $N$, and $S$ are spanned by the eigenfunctions with positive, vanishing, and negative real part of the stability exponents, respectively.
If there are no exponents with zero real part, the behavior of the nonlinear system, Eq.~(\ref{eq:sys}), is well described close to the equilibrium by the linear system of Eq.~(\ref{eq:sys_lin}).
A more precise description is needed if such exponents exist.
Then $N$, called \emph{center subspace}, is not empty and a so-called \emph{local center manifold} $\mathcal{W}_{\text{loc}}$ exists,
which is invariant in the sense that solutions generated from initial functions on the center manifold evolve on the center manifold.
$\mathcal{W}_{\text{loc}}$ is tangential to the center subspace $N$ and therefore can be expressed by its deviation $h$ from the center subspace,
\begin{equation}
\label{eq:center_manifold}
\mathcal{W}_{\text{loc}}\cap V=\{\phi+h(\phi)\;|\;\phi\in N\cap V\},
\end{equation}
where $V$ is a neighborhood of $0\in C$, i.e., a neighborhood of the equilibrium of Eq.~(\ref{eq:sys}) and the function $h$ fulfills $h(0)=0$ as well as $(Dh)(0)=0$.

The dynamics on the center manifold is obtained by the projection of Eq.~(\ref{eq:sys}) onto the center subspace.
To find it we need the basis $\Phi_c$ of the center subspace and the basis $\Psi_c$ of its dual.
Then the projection $x_t^N$ of the solution $x_t$ to the center subspace is given by $x_t^N=\Phi_cy(t)$, where the time-varying coefficients $y$ are given by $y(t)=(\Psi_c,x_t)$
and $(\cdot,\cdot)$ is a suitable bilinear form for this problem such that $\Phi_c$, $\Psi_c$, and the other eigenfunctions of Eq.~(\ref{eq:sys_lin}) form a biorthogonal system with respect to $(\cdot,\cdot)$.
The coefficients $y(t)$ fulfill the ODE
\begin{equation}
\label{eq:y}
\dot{y}=B_cy+\Psi_c(0)f(\Phi_cy+h(\Phi_cy)),
\end{equation}
where the matrix $B_c$ fulfills $T(t)\Phi_c=\Phi_ce^{B_ct}$ and $T(t)$ is the solution operator of the linear system, Eq.~(\ref{eq:sys_lin}).
The initial value $y_0=y(0)$ is given by $y_0=(\Psi_c,\phi)$, where $\phi(t)$ is the initial function for Eq.~(\ref{eq:sys}).
Using the definition of the center manifold, the dynamics on this space is then given by
\begin{equation}
\label{eq:x}
x_t=\Phi_cy(t)+h(\Phi_cy(t)).
\end{equation}

For our example from Eq.~(\ref{eq:sys_our}) the center subspace is one-dimensional with basis $\Phi_c=1$ and its dual $\Psi_c=1/(\Theta+1)$, where the bilinear form reads
\begin{equation}
(\psi,\phi)=\psi(0)\phi(0)+\Theta\int_{-1}^0\psi(t'+1)\phi(t')\,\text{d}t'.
\end{equation}
Here, we modified the eigenfunctions from Ref.~\cite{amann2007} such that they fit into the mathematical framework in Ref.~\cite{hale1993}.
The matrix $B_c$ degenerates to a scalar $B_c=0$, which is equal to the only stability exponent with zero real part.
The ODE for the coefficient $y(t)$ reads
\begin{equation}
\dot{y}=\frac{\Theta}{\Theta+1}4\pi^3\mu[y+h(y)(-1)]^3
\end{equation}
with the initial condition $y_0=(\Psi_c,\phi)=1/(\Theta+1)[\phi(0)+\Theta\int_{-1}^0\phi(t')\,dt']$, where $\phi$ is the initial function for Eq.~(\ref{eq:sys_our}).
Since we have no exact expression for $h$, we have to find an approximation of the ODE, where we use that $h(y)$ is of third order since $(Dh)(0)=0$ and $h$ inherits the reflectional symmetry of the nonlinear DDE,
which is also used in Ref.~\cite{redmond2002} for deriving the dynamics on the center manifold of a similar DDE.
Neglecting all terms with a larger order than three on the right hand side of the ODE leads to
\begin{equation}
\dot{y}=\frac{\Theta}{\Theta+1}4\pi^3\mu y^3.
\end{equation}
Finally, close to the equilibrium one has
\begin{equation}
x_t(t')=\Phi_c(t')y(t)+h(\Phi_c y(t))(t')\approx\Phi_c(t')y(t)=y(t).
\end{equation}

Except for the prefactor, this ODE is identical to the continuous-time approximation of Pomeau-Manneville like iterated maps in the vicinity of marginally unstable fixed points \cite{geisel1984,bel2006}.
From these iterated maps it is known that the distribution of residence times close to the marginally unstable fixed points
possesses the same heavy tail as the one we observe for the laminar phases in our delay system.

\section{\label{sec:4}Modulated Delay: Anomalous Behavior}

\begin{figure*}
\includegraphics[width=\linewidth]{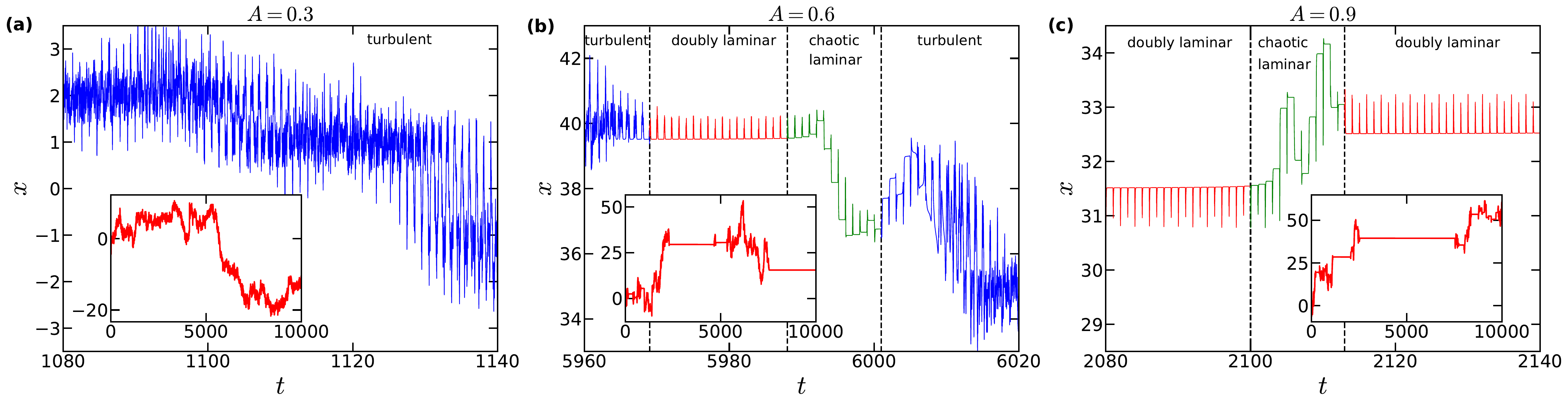}
\caption{\label{fig:3}
Numerical solutions of the DDE (Eqs.~(\ref{eq:DDE}-\ref{eq:f_x})) in the case of a modulated delay for three different values of the modulation amplitude ($A=0.3,0.6,0.9$ from left to right)
show an increase of the fraction of plateau phases (consisting of doubly-laminar and chaotic-laminar phases) as the modulation amplitude is increased.
The main figures show the solutions on a smaller time scale and the insets on a larger time scale ($\Theta=50$, $\tau_0=1$, $\mu=0.9$).}
\end{figure*}

In this section, we consider the more general case of the modulated delay from Eqs.~(\ref{eq:R_t}) and (\ref{eq:tau_t}) with $\tau_0=1$ and $0\leq A<1$.
Our goal is to study the influence of the modulation of the delay time on the occurrence of anomalous chaotic diffusion.
Similar to the case of the constant delay, we first consider some numerical solutions of the DDE, which are shown in Fig.~\ref{fig:3} for different values of the modulation amplitude $A$.
For a smaller value of the modulation amplitude ($A=0.3$), the depicted solution entirely consists of a turbulent phase.
For an intermediate value ($A=0.6$), one recognizes a switching between turbulent phases and plateau phases.
The latter are characterized by a sequence of nearly constant plateaus interrupted by irregular bursts
\footnote{The solution in one state interval is identified as a plateau if the median of all derivatives of the solution inside the state interval is smaller than 0.01.
A plateau belongs to a doubly-laminar phase if the value of the plateau, i.e., the value of the solution taken in the middle of a state interval is closer than 0.05 to the nearest marginally unstable fixed point.
Otherwise, it belongs to a chaotic-laminar phase.}.
For a larger value of the modulation amplitude ($A=0.9$), the shown solution completely consists of a plateau phase.
In order to understand the emergence of these plateau phases, we have to recall the basic principles of the so-called laminar chaos \cite{mueller2018,mueller2019,hart2019,mueller2020}:

The DDE in Eq.~(\ref{eq:DDE}) represents a feedback loop, where a delayed signal is frequency modulated for $A>0$, transformed by the nonlinearity $f(x)$ which,
interpreted as an iterated map $x'=f(x)$, can in general be (weakly) chaotic with Lyapunov exponent $\lambda_f\geq0$, and low-pass filtered with cutoff frequency $\Theta$.
The DDE can show two types of chaos depending on the parameters of the delay:
Conservative delays \cite{otto2017,mueller2017}, where the so-called access map $t'=R(t)$ has zero Lyapunov exponent ($\lambda_R=0$), are equivalent to constant delays ($A=0$, no frequency modulation)
and lead to turbulent chaos with strong fluctuations due to the chaos generating map $x'=f(x)$.
Dissipative delays \cite{otto2017,mueller2017}, where the access map has a negative Lyapunov exponent ($\lambda_R<0$), are not equivalent to constant delays
and the corresponding frequency modulation of the delayed signal can compensate the strong fluctuations generated by the nonlinearity leading to low-dimensional (generalized) laminar chaos
with low-frequency phases (the plateaus in Fig.~\ref{fig:3}).
A necessary condition for laminar chaos is that the stretching of the solution on the time axis due to the retarded argument $R(t)$ is stronger
than the stretching of the solution in coordinate direction due to the nonlinearity $f(x)$.
Expressed with formulas, this means that we must have $\lambda_R+\lambda_f<0$ in this case \cite{mueller2018}.

\begin{figure}
\includegraphics[width=\linewidth]{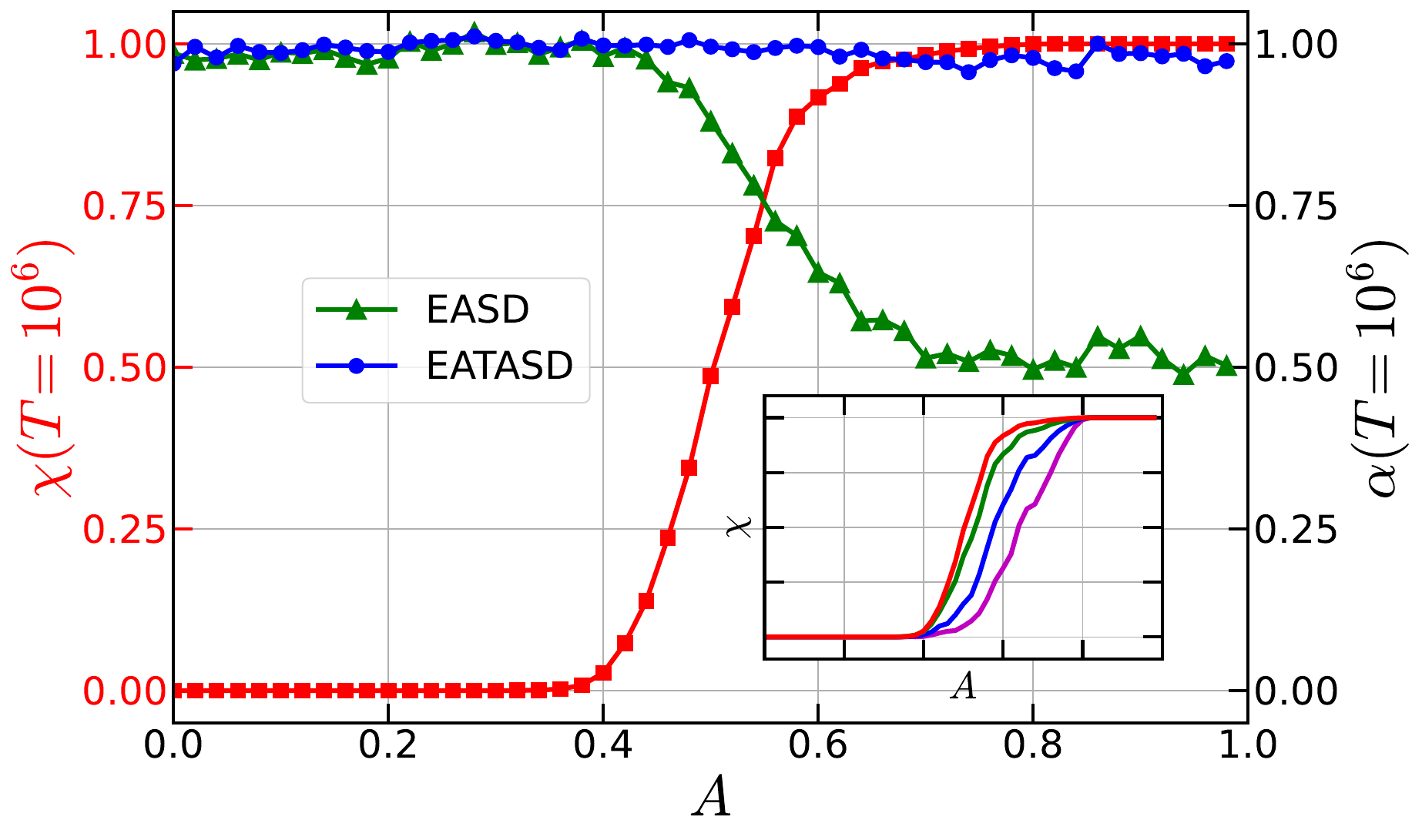}
\caption{\label{fig:4}
The mean fraction $\chi(T)$ of plateau phases (red squares) increases from zero to one under a variation of the amplitude $A$ of the delay modulation
and strongly depends on the observation time $T$ as shown in the inset ($T=10^3,10^4,10^5,10^6$ from bottom to top).
The diffusion exponent $\alpha(T)$ obtained from the EASD (green triangles) shows a corresponding transition:
if the fraction $\chi(T)$ is close to zero, the diffusion exponent $\alpha(T)$ is roughly unity indicating normal diffusion and if $\chi(T)$ is close to one, $\alpha(T)$ is roughly one half indicating subdiffusion.
The diffusion exponent from the EATASD (blue circles) is always roughly equal to unity implying a nonequivalence of ensemble and time averages if $\chi(T)$ deviates from zero (same parameters as in Fig.~\ref{fig:3}).}
\end{figure}

For our modulated delay from Eqs.~(\ref{eq:R_t}) and (\ref{eq:tau_t}) with $\tau_0=1$ and $0\leq A<1$, the Lyapunov exponent $\lambda_R=\ln(1-A)$ depends on the modulation amplitude $A$.
For the Lyapunov exponent $\lambda_f$, the situation is more complex.
The iterated map $x'=f_{\mu}(x)$ corresponding to the double-sine nonlinearity from Eq.~(\ref{eq:f_x}) -- which in the following we refer to as the double-sine map -- shows weak chaos ($\lambda_f=0$)
due to the existence of marginally unstable fixed points $x^*=1/2+n\in\mathbb{Z}$ with $f_{\mu}'(x^*)=1$ at the half-integers.
Therefore, the above mentioned condition for laminar chaos is satisfied for every $A>0$ and one would expect laminar chaos for every $A>0$.
However, according to the numerical solutions of the DDE in Fig.~\ref{fig:3}, this is not the case.
The double-sine map belongs to the class of Pomeau-Maneville maps \cite{manneville1979,pomeau1980,manneville1980},
which are known for the occurrence of intermittency and $1/f$ noise due to very long residence times close to marginally unstable fixed points.
The corresponding distribution $\psi(\delta)$ of these residence times can be inferred from the asymptotic behavior of the map close to the fixed points $x^*$ \cite{geisel1984}
which in our case is given by $f_{\mu}(x)\simeq x+4\pi^3\mu(x-x^*)^3$ resulting in $\psi(\delta)\sim\delta^{-3/2}$.
The mean value of this distribution, i.e., the mean residence time, diverges.
Due to the lack of a finite characteristic time scale, the dynamics becomes nonstationary and shows weak ergodicity breaking.
The latter implies that time-averaged quantities, such as the Lyapunov exponent $\lambda_f$,
remain random variables and follow a broad distribution even in the limit of an infinite measurement time \cite{korabel2009,korabel2010}.
Between the residence times, the dynamics is more chaotic and a corresponding short-time Lyapunov exponent is positive and may violate the condition for laminar chaos.
This may lead to a switching between plateau and turbulent phases as can be seen in the numerical solutions of the DDE in Fig.~\ref{fig:3} (b).
For increasing values of the modulation amplitude $A$, the Lyapunov exponent $\lambda_R=\ln(1-A)$ becomes more negative such that the condition for laminar chaos can be satisfied more easily.
Correspondingly, the mean fraction of plateau phases in the numerical solutions of the DDE increases with increasing modulation amplitude $A$ as can be seen in Fig.~\ref{fig:4}.
Additionally, due to the nonstationarity of the problem, this fraction also depends on the observation time $T$ that is shown in the inset of Fig.~\ref{fig:4}.

As a next step, we want to relate the occurrence of the plateau phases to the appearance of anomalous chaotic diffusion.
In order to characterize diffusion processes in general, one typically considers the asymptotic time dependence of the mean-squared displacement (MSD).
The latter can be defined as an ensemble or a time average.
The ensemble-averaged squared displacement (EASD) $\langle\Delta x^2(t)\rangle_{\text{E}}:=\langle[x(t)-x(0)]^2\rangle_{\text{E}}\sim t^{\alpha}$
typically increases according to a power law with the diffusion exponent $\alpha$.
The time-averaged squared displacement (TASD) is defined by $\langle\Delta x^2(t)\rangle_{\text{T}}:=1/(T-t)\int_0^{T-t}[x(t'+t)-x(t')]^2\,\text{d}t'$ and is a random variable for every finite $T$.
Therefore, we consider its ensemble average (EATASD) $\langle\langle\Delta x^2(t)\rangle_{\text{T}}\rangle_{\text{E}}$.
The diffusion exponents extracted from both kinds of MSDs are shown in Fig.~\ref{fig:4} as a function of the modulation amplitude $A$.
For the EASD, we observe a transition from normal diffusion ($\alpha\approx1$) at low modulation amplitudes to subdiffusion ($\alpha\approx1/2$) for larger modulation amplitudes.
This transition seems to be related to the mean fraction of plateau phases.
From the theory of laminar chaos \cite{mueller2018,mueller2019}, it is known that the heights of the plateaus are mapped from one plateau to the successive one by the action of the nonlinearity $f_{\mu}(x)$.
This means that the dynamics of the plateau heights is determined by the double-sine map.
Correspondingly, plateau phases can be divided into doubly-laminar phases, which are related to the long residence times close to the marginally unstable fixed points,
and chaotic-laminar phases related to the chaotic transitions between these residence times, see Fig.~\ref{fig:3} (b) and (c).
The corresponding distributions of the durations of these phases can be inferred from the map dynamics.
It turns out that while the distribution $\psi_{\text{dl}}(\delta)$ of doubly-laminar phase durations is heavy-tailed with $\psi_{\text{dl}}(\delta)\sim\delta^{-3/2}$,
the distribution $\psi_{\text{cl}}(\delta)$ of chaotic-laminar phase durations decays exponentially.
The dynamics of the double-sine map can be considered on unit cells because of the discrete translational symmetry $f_{\mu}(x\pm1)=f_{\mu}(x)\pm1$ of the underlying nonlinearity.
Chaotic diffusion occurs for the double-sine map because each unit cell contains two ranges that map to the left and right adjacent cell, respectively,
with equal probability due to the reflection symmetry $f_{\mu}(-x)=-f_{\mu}(x)$ \cite{geisel1982}.
Moreover, the chaotic diffusion becomes subdiffusive due to the long residence times close to the marginally unstable fixed points in the center of each cell \cite{geisel1984}.
This dynamical behavior can be modeled by a subdiffusive continuous-time random walk (CTRW) with instantaneous jumps and waiting times
that are randomly drawn from a distribution $\psi(\delta)\sim\delta^{-\alpha-1}$ with a heavy tail \cite{zumofen1993}.
In this model, the diffusion exponent obtained from the EASD is equal to $\alpha$, whereas the diffusion exponent determined from the EATASD is always equal to unity \cite{lubelski2008,he2008}.
This different scaling behavior of ensemble and time averages although the full state space is accessible is typically referred to as weak ergodicity breaking \cite{bouchaud1992}.
The dynamics of the double-sine map is reflected in the delay system in the case of laminar chaos.
For small values of the modulation amplitude $A$, the mean fraction of plateau phases is equal to zero indicating that only turbulent chaos is present, which is known to lead to normal diffusion \cite{albers2022_1}.
For large values of the modulation amplitude $A$, the mean fraction of plateau phases is close to unity indicating that laminar chaos prevails.
Because the dynamics of the heights of the plateaus is governed by the double-sine map, one finds subdiffusion with $\alpha\approx1/2$.
For intermediate values of the modulation amplitude $A$, one observes a transition from normal to anomalous subdiffusion
corresponding to the transition of the mean fraction of plateau phases from pure turbulent chaos to pure laminar chaos.
The diffusion exponents obtained from the EATASD are all equal to one, demonstrating that also the weak nonergodicity of the double-sine map is reflected in the delay system.

\begin{figure*}
\includegraphics[width=\linewidth]{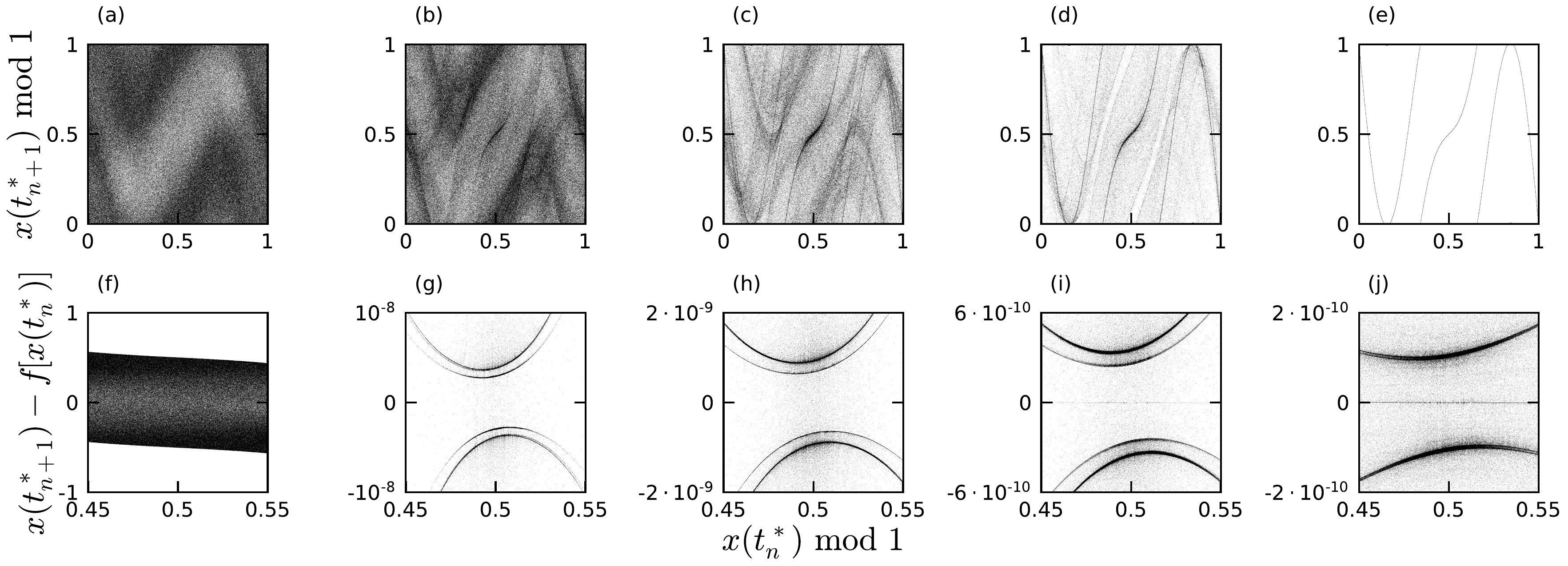}
\caption{\label{fig:5}
Return maps obtained by sampling trajectories in the middle of each state interval (at the attractive fixed points $t_n^*$ of $R(t)+1$)
and plotting the dependence of two successive values $x(t_{n+1}^*)$ and $x(t_n^*)$ for different values of the modulation amplitude $A$ ($A=0.3,0.5,0.6,0.7,0.9$ from left to right).
At first glance, for larger values of $A$ (where laminar chaos prevails), the return maps reproduce the double-sine map especially in the vicinity of the marginally unstable fixed point (upper row),
but on a finer scale by plotting the deviations from this iterated map in the lower row, one recognizes multiple branches (unperturbed and perturbed ones).
For decreasing values of $A$, the return maps become more and more fuzzy (same parameters as in Fig.~\ref{fig:3}).}
\end{figure*}

The dependence of the fraction of plateau phases on the modulation amplitude $A$ is also related to the return maps depicted in Fig.~\ref{fig:5}.
These maps show the dependence of the solution value in the middle of a state interval on the corresponding value in the previous interval.
For large values of $A$, where laminar chaos prevails, the return maps reproduce the graph of the double-sine map because the latter determines the dynamics of the plateau heights.
For decreasing values of $A$, this graph is increasingly blurred.
The reason for the blurring is the increasing fraction of turbulent phases between the plateau phases which cannot be described by a simple one-dimensional iterated map.

\begin{figure}
\includegraphics[width=\linewidth]{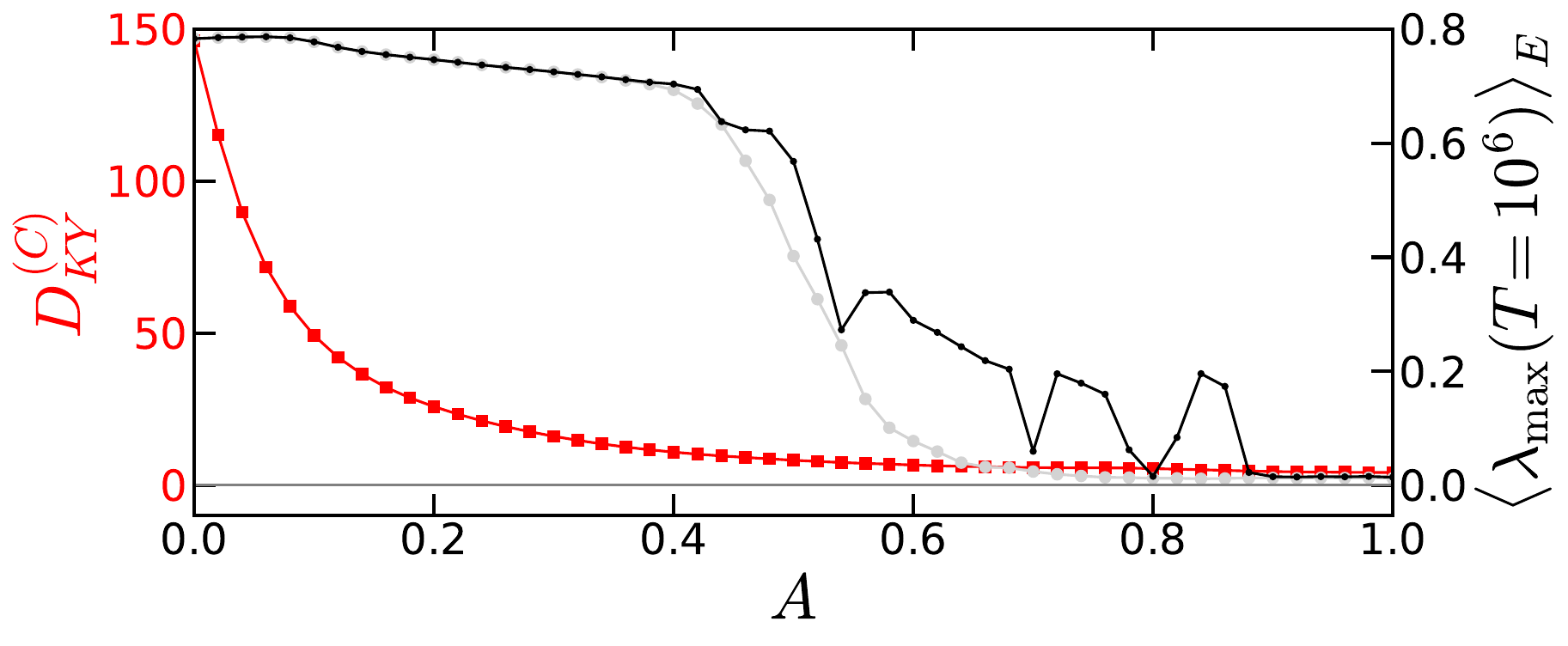}
\caption{\label{fig:6}
The effective dimension of the chaotic phases (turbulent and chaotic laminar) as measured by their Kaplan-Yorke dimension $D_{\text{KY}}^{(\text{C})}$ (red squares)
decreases quickly to zero for increasing values of the modulation amplitude $A$.
The ensemble-averaged maximal finite-time Lyapunov exponent $\langle\lambda_{\text{max}}(T)\rangle_{\text{E}}$ (small black dots)
shows a similar transition as the mean fraction of plateau phases shown in Fig.~\ref{fig:4} under a variation of $A$.
A comparison with the contribution of the chaotic phases to the maximal finite-time Lyapunov exponent (grey dots) reveals that the irregular behavior for intermediate $A$
is caused by the contribution from the doubly-laminar phases (see text).}
\end{figure}

We are also interested in the fractal dimension of the chaotic phases between the doubly-laminar phases, which can be measured by their Kaplan-Yorke dimension (see Fig~\ref{fig:6})
\footnote{The Kaplan-Yorke dimension can be directly calculated from the Lyapunov spectrum, i.e., the set of all Lyapunov exponents.
The latter were defined as time averages of the expansion rates inside only the chaotic-laminar and turbulent phases.
They were numerically calculated by using the method in \cite{farmer1982} and in order to supress fluctuations, the obtained exponents were averaged over 1024 initial functions.}.
The decaying dimension of the chaotic phases increases their probability of being captured by the nonhyperbolic fixed-point solutions for increasing values of $A$.

The increasing fraction of plateau phases is also reflected in the maximal finite-time Lyapunov exponent of the delay system, which goes to zero as the modulation amplitude increases (see Fig.~\ref{fig:6}).
As a consequence, the delay system is weakly chaotic for large values of $A$.
However, the approach to zero is very irregular, which will be discussed in more detail in the next section.

\section{\label{sec:5}Irregular behavior of the Lyapunov exponent and burst dynamics}

\begin{figure}
\includegraphics[width=\linewidth]{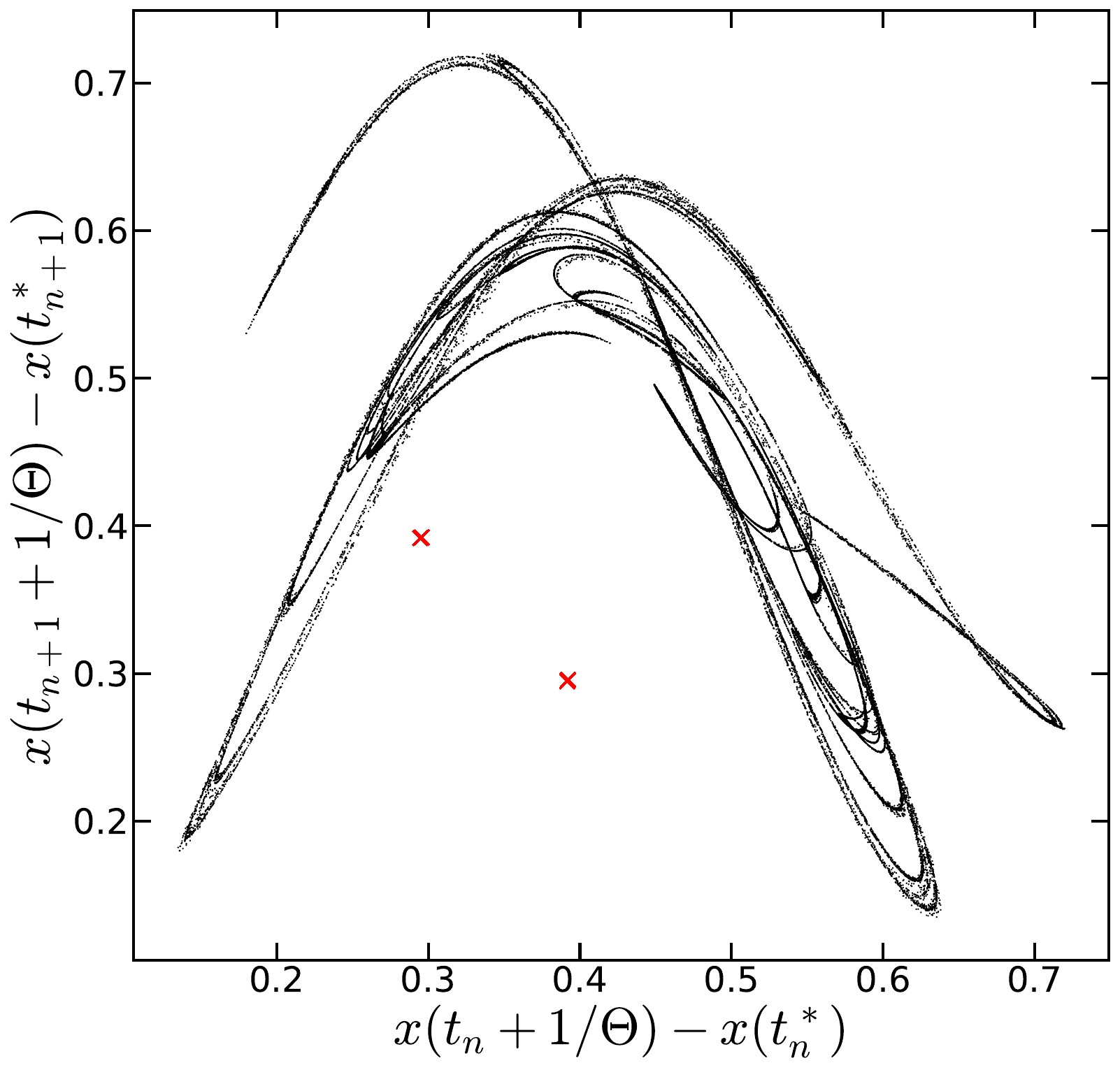}
\caption{\label{fig:7}
Coexistence of different burst dynamics:
Return maps of snapshots of the bursts between the plateaus for two doubly-laminar phases show chaotic (black dots, max. expansion rate $\lambda_k^{(L)}\approx0.261$, Kaplan-Yorke dimension: $D_{\text{KY},k}^{(L)}\approx2.63$) and periodic (red crosses, $\lambda_{k'}^{(L)}\approx0.002$, $D_{\text{KY},k'}^{(L)}\approx1.01$) burst dynamics.
We set $A=0.74$, where the return maps were generated from 50000 (chaotic) and 1000 (periodic) bursts and the other quantities were computed from the whole doubly-laminar phase.}
\end{figure}

In order to understand the irregular behavior of the ensemble-averaged maximal finite-time Lyapunov exponent $\langle\lambda_{\mathrm{max}}(T)\rangle_E$,
we express it as $\lambda_{\text{max}}(T)=\chi^{(C)}(T)\lambda^{(C)}(T)+(1-\chi^{(C)}(T))\lambda^{(L)}(T)$,
where $\lambda^{(C)}(T)$ and $\lambda^{(L)}(T)$ are the Lyapunov exponents of the chaotic (turbulent and chaotic-laminar) and doubly-laminar phases, respectively,
and $\chi^{(C)}$ is the fraction of the chaotic phases.
In Fig.~\ref{fig:6}, the contribution $\langle\chi^{(C)}(T)\lambda^{(C)}(T)\rangle\approx\langle\chi^{(C)}(T)\rangle\langle\lambda^{(C)}(T)\rangle$ of only the chaotic phases (grey dots)
to the ensemble-averaged maximal finite-time Lyapunov exponent $\langle\lambda_{\mathrm{max}}(T)\rangle_E$ (black dots) is shown as a function of $A$.
In contrast to the overall Lyapunov exponent, the curve is very regular, which leads to the conclusion that the irregular behavior is caused by the doubly-laminar phases.
In detail, the irregular behavior of the Lyapunov exponent for intermediate values of $A$ can be explained by the dynamics of the bursts in the doubly-laminar phases,
where stable (nearly) periodic and chaotic burst dynamics is observed.
These two types of dynamics are illustrated in Fig.~\ref{fig:7}, where return maps for two doubly-laminar phases of one trajectory for a fixed parameter set are shown.
The return maps are generated from snapshots of the bursts as follows.
From the theory of laminar chaos \cite{mueller2018,mueller2019}, we know that the bursts are located at the unstable fixed points $t_n$ of $R(t)+1$.
The value of the solution segment $x_n(t)$ with $t=t_n^*$ is the height of the plateau before the burst at $t_n$, where the $t_n^*$ are the stable fixed points of $R(t)+1$.
Due to the asymmetric kernel of the solution operator of Eq.~(\ref{eq:mos}) with width $1/\Theta$, the bursts begin at the $t_n$ and are of a width of the order of $1/\Theta$.
Therefore, snapshots of the bursts can be generated by sampling the time series $x(t)$ at $t=t_n+1/\Theta$.
Subtracting the plateau values $x(t_n^*)$ removes the drift inside the doubly-laminar phases and the height difference between different doubly-laminar phases,
which allows the computation of bounded return maps and a comparison of the burst dynamics of different doubly-laminar phases, respectively.

\begin{figure}
\includegraphics[width=\linewidth]{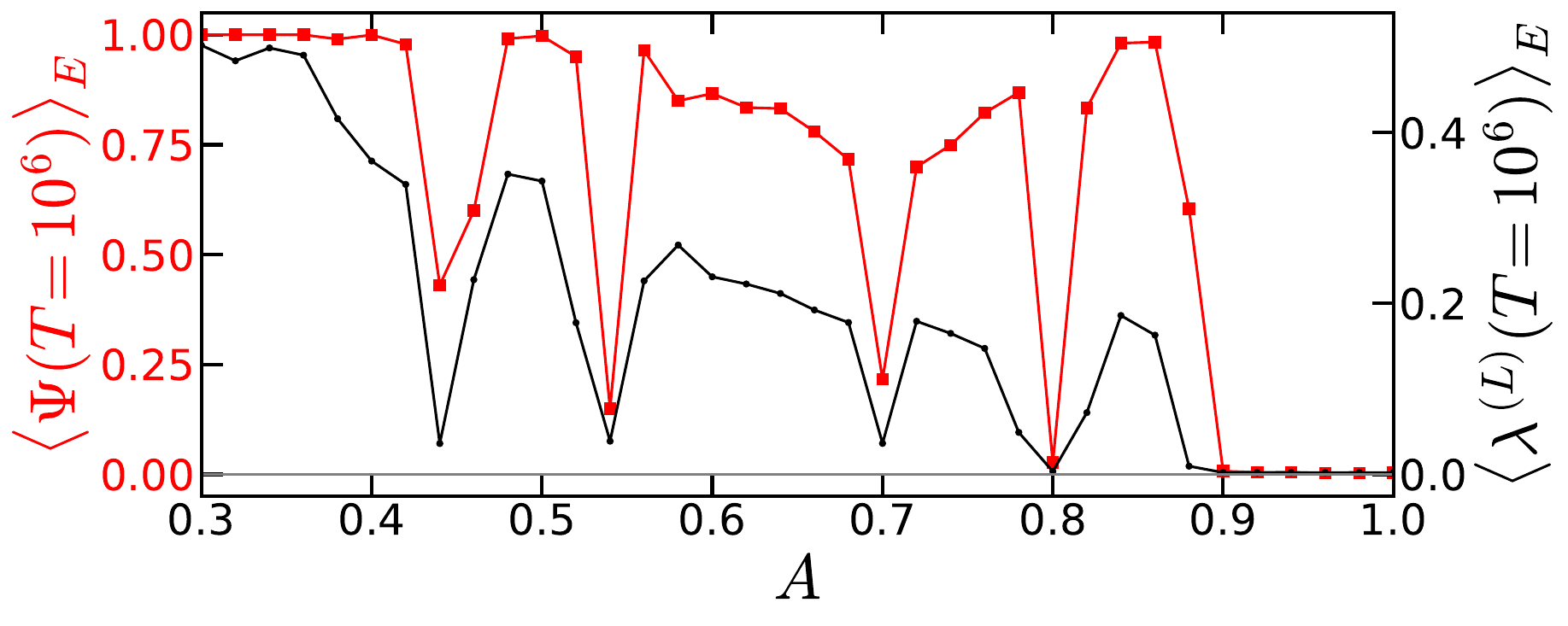}
\caption{\label{fig:8}
Chaotic burst dynamics lead to positive contributions to the maximum Lyapunov exponent:
The maximum finite-time Lyapunov exponent of the doubly-laminar phases (black dots) shows positive correlation with the fraction of bursts with chaotic dynamics (red squares).
In the excluded $A$ range, no doubly-laminar phases were observed.}
\end{figure}

We analyze the influence of the burst dynamics via two quantities.
The influence of a single doubly-laminar phase can be quantified by the expansion rate $\lambda_k^{(L)}=\gamma_k^{(L)}/\delta_k^{(L)}$,
which is the growth rate from the growth exponent $\gamma_k^{(L)}$ of the dominating Lyapunov vector $x_k(t)$ inside the $k$th doubly-laminar phase with duration $\delta_k^{(L)}$.
As illustrated in Fig.~\ref{fig:7} together with values $\lambda_k^{(L)}$ given in the caption, chaotic burst dynamics is associated with positive expansion rates $\lambda_k^{(L)}$
and thus increases the overall Lyapunov exponent in contrast to stable periodic burst dynamics, where we have $\lambda_k^{(L)}\approx0$.
The second quantity is motivated by the fact that chaotic and periodic burst dynamics both can appear within a single time series
so that the burst dynamics and the expansion rate $\lambda_k^{(L)}$ change from doubly-laminar phase to doubly-laminar phase.
We therefore consider the fraction $\Psi(T)$ of chaotic burst dynamics and write the finite-time Lyapunov exponent $\lambda^{(L)}(T)$ of the doubly-laminar phases in a time series of length $T$ as
\begin{equation}
\begin{split}
\lambda^{(L)}(T)=&\Psi(T)\frac{1}{\Delta^{(L,\text{cb})}(T)}\sum_{j=0}^{J(T)-1}\gamma^{(L,\text{cb})}_j\\
&+(1-\Psi(T))\frac{1}{\Delta^{(L,\text{pb})}(T)}\sum_{m=0}^{M(T)-1}\gamma^{(L,\text{pb})}_m,
\end{split}
\end{equation}
where $\Delta^{(L,\text{cb})}(T)$ ($\Delta^{(L,\text{pb})}(T)$) and $\gamma^{(L,\text{cb})}_m$ ($\gamma^{(L,\text{pb})}_j$), respectively,
are the sum of all durations of doubly-laminar phases and the growth exponents of the dominating Lyapunov vector inside the doubly-laminar phases with chaotic (periodic) burst dynamics.
The fraction $\Psi(T)$ is then given by $\Psi(T)=\Delta^{(L,\text{cb})}(T)/[\Delta^{(L,\text{cb})}(T)+\Delta^{(L,\text{pb})}(T)]$.
To classify the burst dynamics, we computed the expansion rates of the first three dominating Lyapunov vectors inside each doubly-laminar phase
of an ensemble of $1024$ time series of length $T=10^6$ using a generalized version of the method discussed in Ref.~\cite{farmer1982} with an integration step size of $\Delta t=0.001$.
Considering these expansion rates as finite-time Lyapunov spectrum inside the $k$th doubly-laminar phase, we computed the Kaplan-Yorke dimension $D_{\text{KY},k}^{(L)}$ for each phase
and classified the burst dynamics as chaotic if we have $D_{\text{KY},k}^{(L)}>2$, otherwise periodic burst dynamics is assumed.
The resulting ensemble average $\langle\Psi(T)\rangle_E$ of the fraction of chaotic burst dynamics is shown together with the ensemble-averaged finite-time Lyapunov exponent $\langle\lambda^{(L)}(T)\rangle_E$
of the doubly-laminar phases as a function of $A$ in Fig.~\ref{fig:8}, where $\lambda^{(L)}(T)$ determines the contribution of these phases to the overall maximum Lyapunov exponent.
We find a strong positive correlation indicating that the irregular behavior of the burst dynamics under variation of $A$
causes the irregular behavior of $\langle\lambda_{\mathrm{max}}(T)\rangle_E$ observed in Fig.~\ref{fig:6}.

\section{\label{sec:6}Anomalous behavior for large modulation amplitudes}

\begin{figure}
\includegraphics[width=\linewidth]{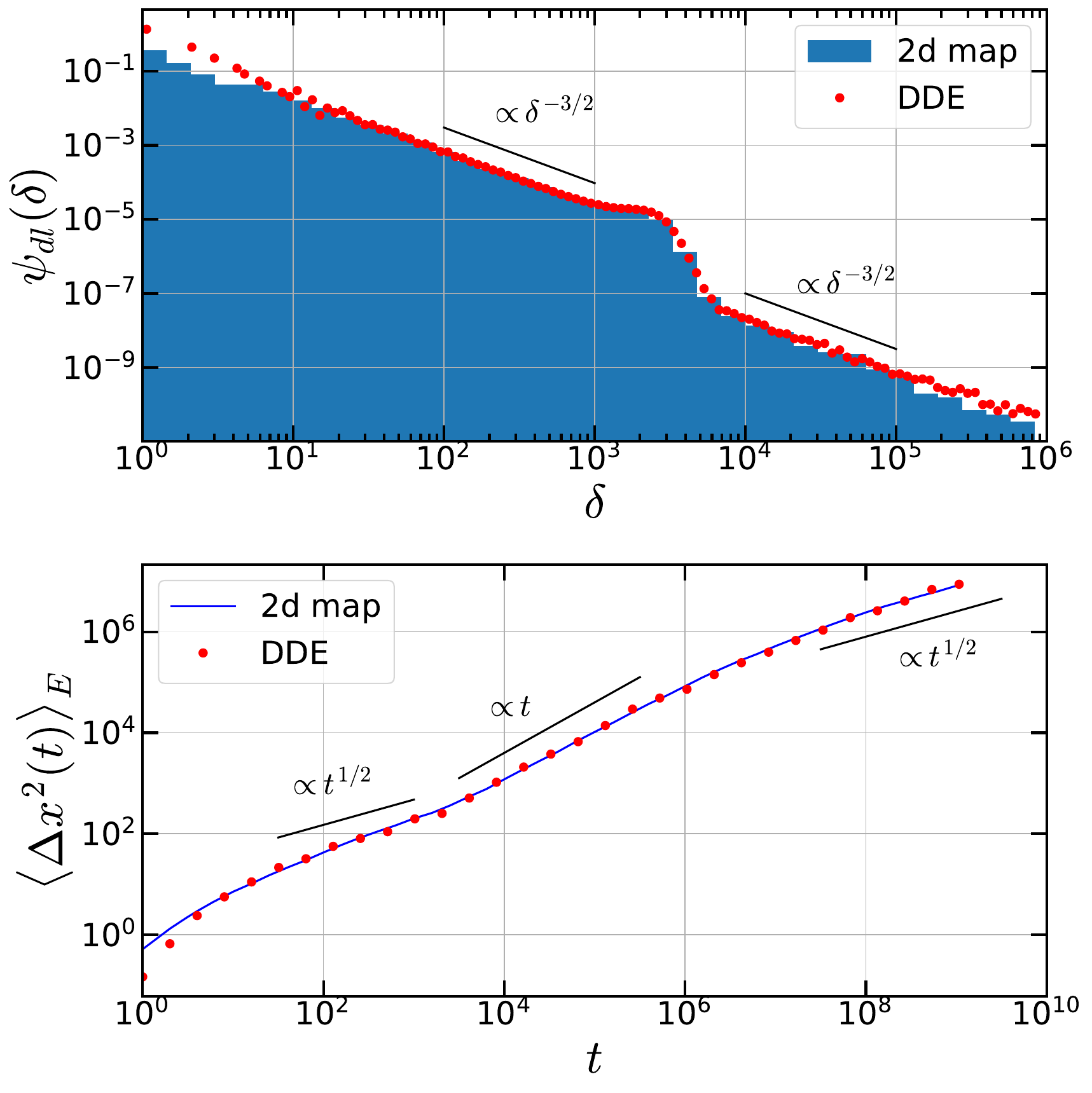}
\caption{\label{fig:9}
For larger values of $A$ (here $A=0.98$), the distribution $\psi_{\text{dl}}(\delta)$ of doubly-laminar phases (red circles, upper part)
possesses an exponential transition from a preasymptotic heavy tail to an asymptotic one with the same exponent.
Correspondingly, the EASD $\langle\Delta x^2(t)\rangle_{\text{E}}$ (red circles, lower part) shows a transition from subdiffusion to normal diffusion and finally to asymptotic subdiffusion.
These numerical results obtained from the DDE (Eqs.~(\ref{eq:DDE}-\ref{eq:f_x})) can be very well reproduced by the two-dimensional iterated map from Eq.~(\ref{eq:2d_map}).}
\end{figure}

The appearance of anomalous chaotic diffusion in our delay system is partially more complex than one would naively expect.
For larger values of the modulation amplitude $A$, the distribution $\psi_{\text{dl}}(\delta)$ of doubly-laminar phases consists of a preasymptotic heavy tail with an exponential cutoff
and an asymptotic heavy tail with the same exponent (see Fig.~\ref{fig:9} top).
This distribution determines the time dependence of the EASD.
The latter first increases with a power law related to the preasymptotic heavy tail in $\psi_{\text{dl}}(\delta)$ followed by a linear increase due to the exponential cutoff of this heavy tail.
Asymptotically, one finds subdiffusion related to the asymptotic heavy tail in the distribution of doubly-laminar phases (see Fig.~\ref{fig:9} bottom).
This behavior can be explained by the return maps in Fig.~\ref{fig:5}, which visualize the dependence of the solution value in the middle of a state interval on the corresponding value in the previous interval.
At first glance, the return map in (e) reproduces the double-sine map $x'=f_{\mu}(x)$,
but if one considers the deviations from this map in the vicinity of the marginally unstable fixed point, see Fig.~\ref{fig:5} (j), one recognizes essentially three branches (one unperturbed and two perturbed ones).
For the perturbed branches, the plateaus are mapped by a slightly perturbed double-sine map $x'=f_{\mu}(x)\pm\varepsilon$
which is responsible for the preasymptotic heavy tail in the distribution $\psi_{\text{dl}}(\delta)$ that shows an exponential cutoff.
For the unperturbed branch, the plateaus follow exactly the double-sine map, which is responsible for the asymptotic heavy tail and, therefore, for the asymptotic anomalous diffusion.
In the next section, we will show that this kind of dynamical behavior can be reproduced by a two-dimensional iterated map.
Furthermore, an appropriate stochastic model consisting of exponentially distributed phases of normal diffusion interrupted by waiting times distributed according to $\psi_{\text{dl}}(\delta)$
is also able to reproduce these results and will be discussed in the section after next.
Additionally, in the Supplemental Material, we provide an animation, which visualizes this kind of dynamics.

\section{\label{sec:7}Iterated Map}

\begin{figure*}
\centerline{\includegraphics[width=1.2\linewidth]{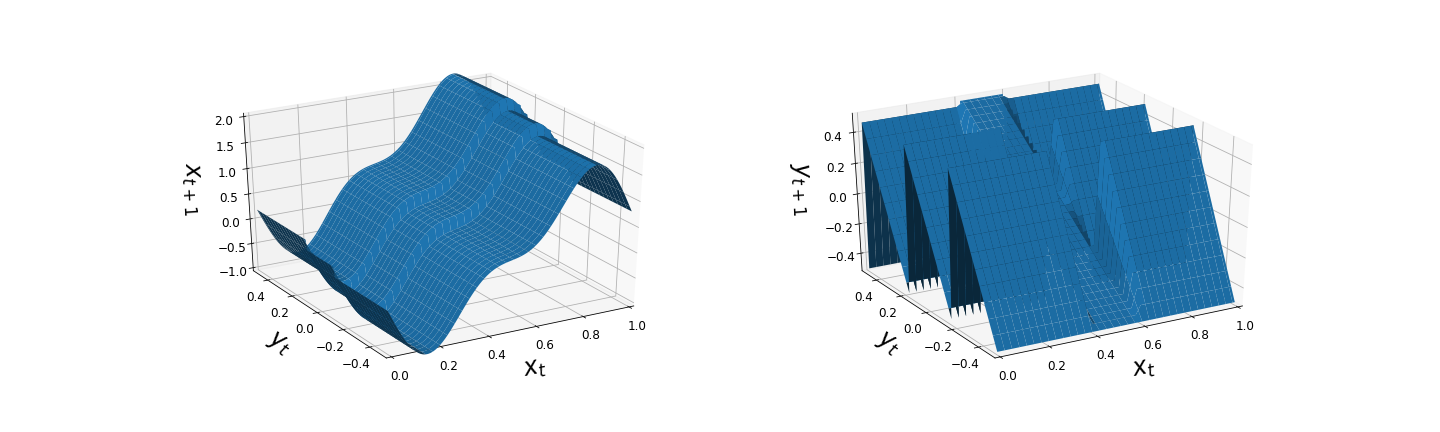}}
\caption{\label{fig:10}
The left figure shows that the first part of the two-dimensional iterated map, Eq.~(\ref{eq:2d_map}), effectively consists of three one-dimensional mappings,
namely the pure double-sine map $x_{t+1}=f_{\mu}(x_t)$ and two slightly disturbed double-sine maps $x_{t+1}=f_{\mu}(x_t)\pm\varepsilon$ depending on the value of the second variable $y_t$.
The dynamics of $y_t$ is given by the second part of the two-dimenional iterated map shown in the right figure.
It is the identity $y_{t+1}=y_t$ if $\tilde{x}_t$ is close to the marginally unstable fixed point of the double-sine map, i.e., ($0.4\leq\tilde{x}_t\leq 0.6$),
therefore fixing the one-dimensional mapping in the left figure during the laminar phases.
Outside the vicinity of this fixed point, $y_t$ follows a chaotic dynamics, namely a modified Bernoulli shift $y_{t+1}=3(y_t+0.5)\text{ mod }1-0.5$, leading to an alternation between the one-dimensional mappings.
The Bernoulli shift has a uniform invariant density on the interval $[-0.5,0.5]$, and therefore, the parameter $p$ in Eq.~(\ref{eq:2d_map}) sets the probability for the iteration of the unperturbed double-sine map.}
\end{figure*}

In order to obtain a deeper understanding of the numerical results from Fig.~\ref{fig:9}, we introduce a purely deterministic, two-dimensional iterated map,
which captures the essential ingredients of the underlying DDE dynamics and is defined by
\begin{equation}
\begin{split}
\label{eq:2d_map}
x_{t+1}&=f_{\mu}(x_t)+\boldsymbol{(}\Theta(y_t-p/2)+\Theta(y_t+p/2)-1\boldsymbol{)}\cdot\varepsilon\\[1ex]
y_{t+1}&=y_t+\boldsymbol{(}\Theta(0.4-\tilde{x}_t)+\Theta(\tilde{x}_t-0.6)\boldsymbol{)}\times\\
&\hspace{5em}\times\boldsymbol{(}3(y_t+0.5)\text{ mod }1-0.5-y_t\boldsymbol{)},
\end{split}
\end{equation}
where $f_{\mu}(x)$ is the double-sine nonlinearity from Eq.~(\ref{eq:f_x}), $\tilde{x}_t$ denotes the corresponding coordinate variable restricted to the unit cell, i.e., $\tilde{x}_t=x_t\text{ mod }1$,
and $\Theta(x)$ is the standard Heaviside step function.
This two-dimensional map is visualized in Fig.~\ref{fig:10}, where for the sake of a better illustration the parameters are partially set to different values than the ones later used for a comparison with the DDE
($\mu=0.9$, $p=0.2$, and $\varepsilon=0.2$).

The two-dimensional iterated map takes into account the dynamics of the pure double-sine map $x_{t+1}=f_{\mu}(x_t)$ and the two slightly perturbed double-sine maps $x_{t+1}=f_{\mu}(x_t)\pm\varepsilon$
as well as the chaotic switching between these maps via the auxiliary variable $y_t$ (see the caption of Fig.~\ref{fig:10} for a detailed explanation of the map).
In order to compare numerical results obtained from this map with the numerical results of the DDE, we have to determine the parameters $p$ and $\varepsilon$.
The parameter $\varepsilon=2\cdot10^{-6}$ can directly be inferred from the corresponding return map.
In order to find the parameter $p$, we approximate the distribution $\psi_{\text{dl}}(\delta)$ from Fig.~\ref{fig:9} by two power laws.
While the preasymptotic heavy tail can be approximated by $0.68\times\delta^{-3/2}$, the asymptotic heavy tail is well described by $0.02\times\delta^{-3/2}$.
The relative weight of the asymptotic heavy tail determines the parameter $p\approx0.02/(0.68+0.02)\approx0.028$.
With these parameter values, the two-dimensional map can be iterated in order to generate trajectories.
We are interested in the distribution $\psi(\delta)$ of the durations of laminar phases, defined as the number of successive iterations inside the vicinity of the marginally unstable fixed points,
and the mean-squared displacement.
For the parameter choice $\mu=0.9$, $p=0.028$, and $\varepsilon=2\cdot10^{-6}$, we obtain the numerical results shown in Fig.~\ref{fig:9}.
We can see a very good agreement with the numerically determined data from the corresponding DDE.
Therefore, it turns out that the dynamics of the unperturbed and perturbed double-sine maps with the correct relative weight are sufficient to explain the observed behavior from the previous section.

\section{\label{sec:8}Stochastic Model}

\begin{figure}
\includegraphics[width=\linewidth]{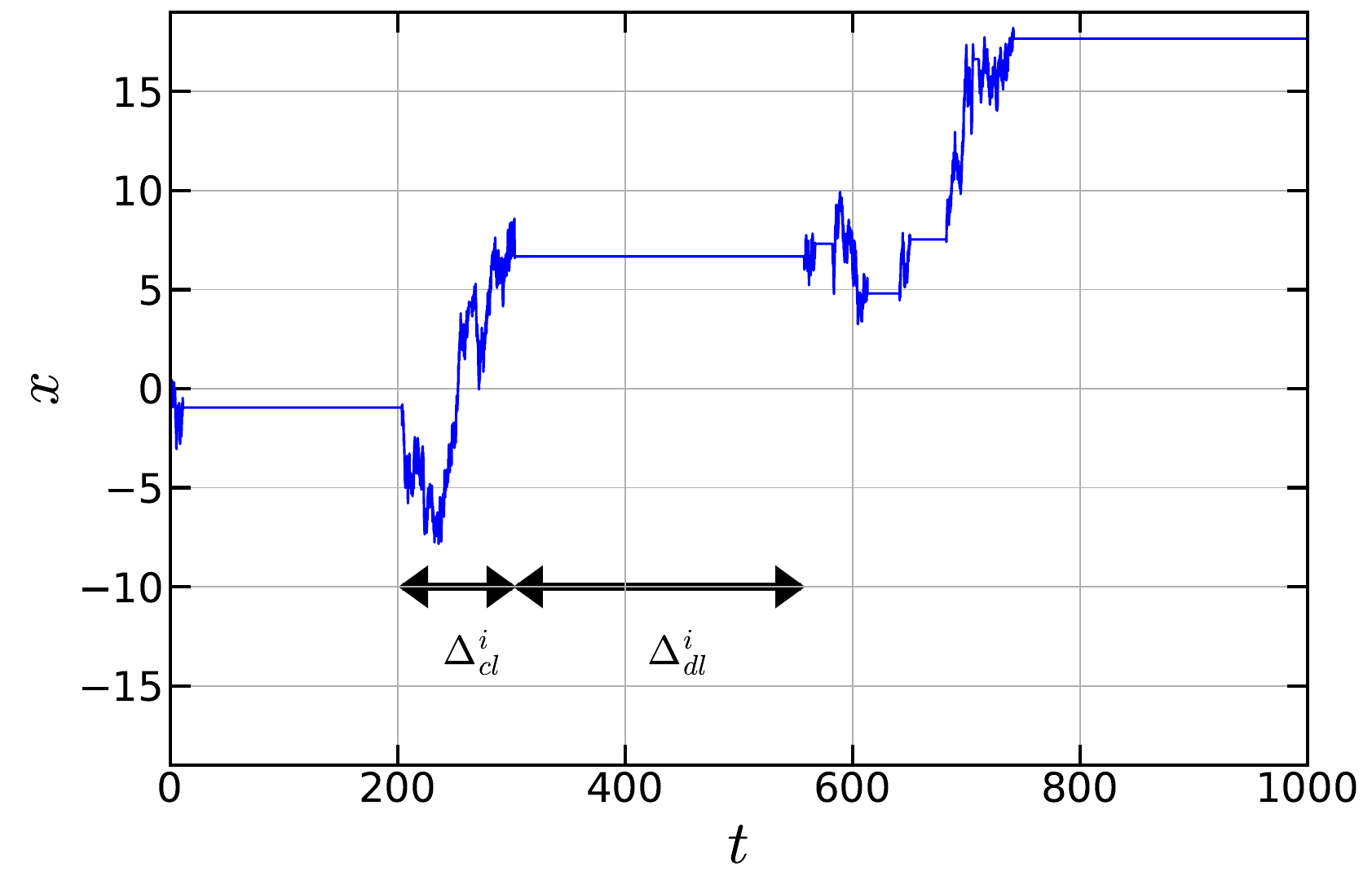}
\caption{\label{fig:11}
The stochastic model is a two-state stochastic process alternating between sequences of Wiener processes,
whose random durations $\Delta_{\text{cl}}^i$ are exponentially distributed according to $\psi_{\text{cl}}(\delta)$,
and random waiting times $\Delta_{\text{dl}}^i$ following the asymptotically heavy-tailed probability density $\psi_{\text{dl}}(\delta)$ described in the main text.}
\end{figure}

The purpose of the following section is to demonstrate that the distributions of the chaotic-laminar phases and the doubly-laminar phases in the solutions of the DDE
determine the time dependence of the mean-squared displacement (MSD).
In order to do so, we consider a stochastic model which consists of a Wiener process which is interrupted at random times by waiting times of random duration (see Fig.~\ref{fig:11}).
As in the well-known continuous-time random walk (CTRW) model, the process starts with a random waiting time $\Delta_{\text{dl}}^1$.
The subscript ``dl'' indicates that the waiting times shall model the doubly-laminar phases.
In contrast to the CTRW, the random waiting time is not followed by an instantaneous jump but by a Wiener process of random duration $\Delta_{\text{cl}}^1$, which models a chaotic-laminar ``cl'' phase.
The process is then renewed.
We assume that the random durations $\Delta_{\text{cl}}^i$ and $\Delta_{\text{dl}}^i$ are independent and identically distributed random variables
drawn from the probability distributions $\psi_{\text{cl}}(\delta)$ and $\psi_{\text{dl}}(\delta)$, respectively.
We first characterize the Wiener process and make assumptions on the involved probability distributions and then analytically derive the resulting MSD.

We denote by $\psi_{\text{cl}}(x,t)$ the probability density for the occurrence of a Wiener process of duration $t$ and of spatial distance $x$ between the start point and the end point of the process,
\begin{equation}
\label{eq:psi_cl_x_t}
\psi_{\text{cl}}(x,t)=W(x|t)\,\psi_{\text{cl}}(t).
\end{equation}
Here, $W(x|t)$ is the propagator of the Wiener process,
\begin{equation}
\label{eq:W_x_t}
W(x|t)=\frac{1}{\sqrt{2\pi Dt}}\exp\left(-\frac{x^2}{2Dt}\right)
\end{equation}
with diffusion coefficient $D$ and we assume that the random duration of the Wiener process is drawn from an exponential distribution,
\begin{equation}
\label{eq:psi_cl_t}
\psi_{\text{cl}}(\delta)=\lambda\,e^{-\lambda\delta}.
\end{equation}
For the analytical treatment, we further need the probability $\Psi_{\text{cl}}(x,t)$ to cover a spatial distance $x$ in time $t$ with a Wiener process that is longer than $t$,
\begin{equation}
\label{eq:Psi_cl_x_t}
\Psi_{\text{cl}}(x,t)=W(x|t)\int_t^{\infty}\psi_{\text{cl}}(\delta)\,\text{d}\delta=\frac{1}{\lambda}\psi_{\text{cl}}(x,t).
\end{equation}
In order to model the doubly-laminar phases, we assume that the corresponding distribution $\psi_{\text{dl}}(t)$ can be written as a superposition
\begin{equation}
\label{eq:psi_dl_t}
\psi_{\text{dl}}(t)=w_{\epsilon}\psi_{\epsilon}(t)+w_0\psi_0(t)
\end{equation}
of a heavy-tailed distribution $\psi_{\epsilon}(t)$ with exponential cutoff and a pure heavy-tailed distribution $\psi_0(t)$,
\begin{equation}
\label{eq:psi_0_t}
\psi_0(t)=\gamma\,(t+1)^{-\gamma-1}.
\end{equation}
The probability $\Psi_{\text{dl}}(t)$ for the occurence of a waiting time longer that $t$ is determined by
\begin{equation}
\label{eq:Psi_dl_t}
\Psi_{\text{dl}}(t)=\int_t^{\infty}\psi_{\text{dl}}(\delta)\,\text{d}\delta.
\end{equation}

With these definitions, we can find an analytical expression for the propagator $p(x,t)$, the probability density to be at spatial position $x$ at time $t$.
From investigations of similar two-state models in the literature such as L\'evy walks with rests \cite{zumofen1995} or a two-state model alternating between a CTRW and a L\'evy walk \cite{liu2022},
we obtain the Fourier and Laplace transform of the propagator,
\begin{equation}
\label{eq:p_k_s}
p(k,s)=\frac{\Psi_{\text{dl}}(s)+\psi_{\text{dl}}(s)\,\Psi_{\text{cl}}(k,s)}{1-\psi_{\text{dl}}(s)\,\psi_{\text{cl}}(k,s)}.
\end{equation}
Here, we apply the widely used notation that the variables $(x,t)$ or $(k,s)$ indicate the space we are working in.
The Laplace transform of the MSD can then be obtained by differentiation with respect to $k$ because the Fourier transform of the propagator is a moment-generating function,
\begin{equation}
\begin{split}
\label{eq:MSD}
\langle\Delta x^2(s)\rangle_{\text{E}}&=\int_{\mathbb{R}}x^2\,p(x,s)\,\text{d}x=-\left.\frac{\partial^2}{\partial k^2}p(k,s)\right|_{k=0}\\[1ex]
&=-\frac{\psi_{\text{dl}}(s)\,\Psi_{\text{cl}}''(k=0,s)}{1-\psi_{\text{dl}}(s)\,\psi_{\text{cl}}(s)}\\[1ex]
&-\frac{\left[\Psi_{\text{dl}}(s)+\psi_{\text{dl}}(s)\,\Psi_{\text{cl}}(s)\right]\psi_{\text{dl}}(s)\,\psi_{\text{cl}}''(k=0,s)}{\left[1-\psi_{\text{dl}}(s)\,\psi_{\text{cl}}(s)\right]^2}.
\end{split}
\end{equation}
In order to derive the asymptotic long-time behavior of the MSD, we need the small $s$ and $k$ expansion of the involved distributions.
The Laplace transform of the exponential distribution $\psi_{\text{cl}}(t)$ can be easily calculated
\begin{equation}
\label{eq:psi_cl_s}
\psi_{\text{cl}}(s)=\frac{\lambda}{\lambda+s}\simeq1-\frac{s}{\lambda}\quad(s\rightarrow0).
\end{equation}
Similarly, we obtain the Fourier and Laplace transform of the distribution $\psi_{\text{cl}}(x,t)$,
\begin{equation}
\label{eq:psi_cl_k_s}
\psi_{\text{cl}}(k,s)=\frac{\lambda}{\lambda+s+\frac{Dk^2}{2}}\simeq1-\frac{s}{\lambda}-\frac{Dk^2}{2\lambda}\quad(s\rightarrow0,k\rightarrow0).
\end{equation}
For $\Psi_{\text{cl}}(k,s)$, we get
\begin{equation}
\label{eq:Psi_cl_k_s}
\Psi_{\text{cl}}(k,s)=\frac{1}{\lambda+s+\frac{Dk^2}{2}}\simeq\frac{1}{\lambda}\left(1-\frac{s}{\lambda}-\frac{Dk^2}{2\lambda}\right).
\end{equation}
Because the heavy-tailed distribution $\psi_{\epsilon}(t)$ has an exponential cutoff and, therefore, a finite mean value $\langle T_{\epsilon}\rangle$, the small $s$ behavior can be written as
\begin{equation}
\label{eq:psi_eps_s}
\psi_{\epsilon}(s)\simeq1-\langle T_{\epsilon}\rangle s\quad(s\rightarrow0).
\end{equation}
The small $s$ expansion of the pure heavy-tailed distribution $\psi_0(t)$ is given by
\begin{equation}
\label{eq:psi_0_s}
\psi_0(s)\simeq1-\Gamma(1-\gamma)s^{\gamma}-\frac{1}{\gamma-1}s\quad(s\rightarrow0).
\end{equation}
With these results, we obtain the small $s$ behavior of the full distribution $\psi_{\text{dl}}(t)$,
\begin{equation}
\label{eq:psi_dl_s}
\psi_{\text{dl}}(s)\simeq1-w_0\Gamma(1-\gamma)s^{\gamma}-\left(w_{\epsilon}\langle T_{\epsilon}\rangle+\frac{w_0}{\gamma-1}\right)s,
\end{equation}
and for $\Psi_{\text{dl}}(s)$, we get
\begin{equation}
\label{eq:Psi_dl_s}
\Psi_{\text{dl}}(s)=\frac{1-\psi_{\text{dl}}(s)}{s}.
\end{equation}
In the following, we use these asymptotics in Eq.~(\ref{eq:MSD}) in order to obtain the MSD.
An asymptotic analysis shows, that the second term in Eq.~(\ref{eq:MSD}) dominates.
This can easily be seen, for instance, from the denominator, which goes to zero for $s\rightarrow0$ and is squared in the second term but not in the first term.
The second derivative $-\psi_{\text{cl}}''(k=0,s)=D/\lambda$ is equal to the second moment of the distances covered by a single Wiener process sequence.
This formula has an intuitive explanation.
The second moment of the propagator $W(x|t)$ of the Wiener process is equal to $D\cdot t$.
By setting $t$ equal to the mean duration $1/\lambda$ of the Wiener process sequences, one obtains the above result.
For the following determination of the dominant term, we have to be careful whether we are interested in the asymptotic or preasymptotic behavior of the MSD,
\begin{equation}
\label{eq:MSD_s}
\langle\Delta x^2(s)\rangle_{\text{E}}\simeq\frac{D}{\lambda}\begin{cases}\frac{s^{-\gamma-1}}{w_0\Gamma(1-\gamma)},\quad&s\ll s^*\\[1ex]
\frac{s^{-2}}{w_{\epsilon}\langle T_{\epsilon}\rangle+\frac{w_0}{\gamma-1}+\frac{1}{\lambda}},\quad&s\gg s^*\end{cases},
\end{equation}
where
\begin{equation}
\label{eq:s_*}
s^*=\left[\frac{w_0\Gamma(1-\gamma)}{w_{\epsilon}\langle T_{\epsilon}\rangle+w_0/(\gamma-1)+1/\lambda}\right]^{1/(1-\gamma)}.
\end{equation}
Correspondingly, by an inverse Laplace transform of Eq.~(\ref{eq:MSD_s}), we obtain our final result in the time domain,
\begin{equation}
\label{eq:MSD_t}
\langle\Delta x^2(t)\rangle_{\text{E}}\simeq\frac{D}{\lambda}\begin{cases}\frac{t^{\gamma}}{w_0\Gamma(1-\gamma)\Gamma(1+\gamma)},\quad&t\gg t^*\\[1ex]
\frac{t}{w_{\epsilon}\langle T_{\epsilon}\rangle+\frac{w_0}{\gamma-1}+\frac{1}{\lambda}},\quad&t\ll t^*\end{cases},
\end{equation}
where
\begin{equation}
\label{eq:t_*}
t^*=\left[\frac{w_{\epsilon}\langle T_{\epsilon}\rangle+w_0/(\gamma-1)+1/\lambda}{w_0\Gamma(1-\gamma)}\right]^{1/(1-\gamma)}.
\end{equation}

\begin{figure}
\includegraphics[width=\linewidth]{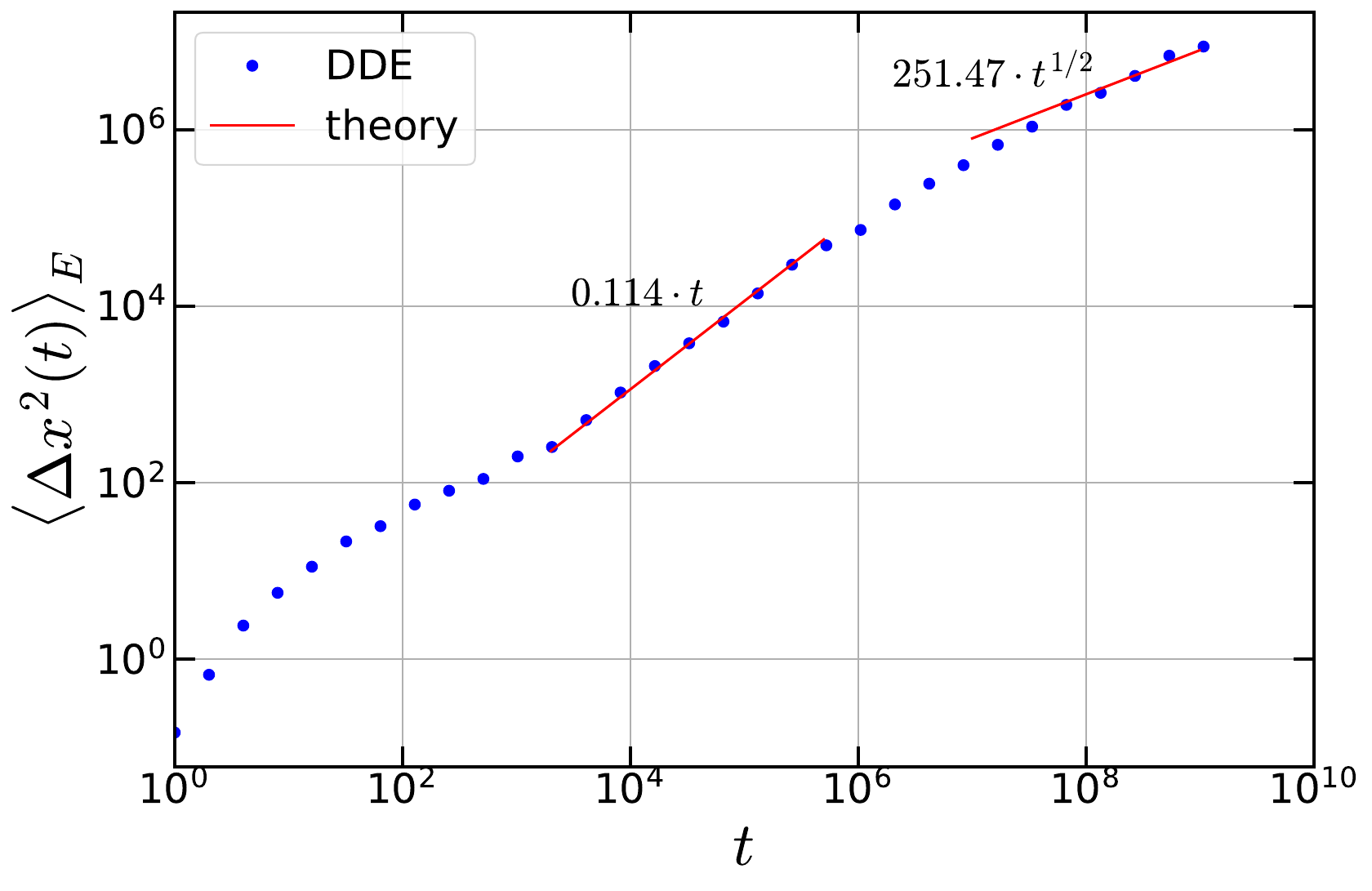}
\caption{\label{fig:12}
The analytic asymptotic subdiffusive and preasymptotic normal diffusive behavior obtained from the MSD of the stochastic model in Eq.~(\ref{eq:MSD_t}) agrees very well with numerical data from the DDE.}
\end{figure}

We now want to compare these formulas with the numerically determined MSD for our DDE in Fig.~\ref{fig:9}.
Specific values for the parameters of the stochastic model can be inferred from the numerically determined distributions of chaotic-laminar phases and doubly-laminar phases,
$\gamma\approx1/2$, $w_{\epsilon}\approx0.96$, $w_0\approx0.04$, $\langle T_{\epsilon}\rangle\approx124.4$, $1/\lambda\approx19.2$, and $D/\lambda\approx15.8$.
A comparison of the analytical results in Eq.~(\ref{eq:MSD_t}) with the MSD obtained from the numerical solutions of our DDE is given in Fig.~\ref{fig:12}.
We can see a very good agreement.
The ``transition point'' $t^*$ between the asymptotic subdiffusion and the preasymptotic normal diffusion can be estimated by Eq.~(\ref{eq:t_*}) resulting in $t^*\approx3.8\cdot10^6$.

As a final point of discussion, we want to mention that the analytical results in Eq.~(\ref{eq:MSD_t}) have interesting physical interpretations.
As indicated in the text above, if one would replace the sequences of Wiener processes by instantaneous jumps, the prefactor $D/\lambda$ could be interpreteted as the second moment of these jumps.
Accordingly, the asymptotic subdiffusive behavior in the first line of Eq.~(\ref{eq:MSD_t}) is exactly the result, one would obtain from the subdiffusive CTRW.
The subdiffusive behavior is caused by the divergence of the time scale of the doubly-laminar phases such that in this case the expontially distributed durations of the chaotic-laminar phases can be neglected.
For the preasymptotic normal diffusive behavior this is, however, not the case.
In this case, we are dealing with two (finite) time scales: the durations of the doubly-laminar phases and the mean duration of the chaotic-laminar phases modeled by the Wiener process.
From the normal random walk theory, it is known that the normal diffusion coefficient is given by the second moment of jumps divided by the mean duration between these jumps \cite{einstein1905}.
The prefactor $D/\lambda$ can be interpreted as the second moment and the denominator in the second line of Eq.~(\ref{eq:MSD_t}) is the sum of the mean durations of chaotic-laminar and doubly-laminar phases,
so in this case, the sequences of Wiener processes cannot be replaced by instantaneous jumps.

\section{\label{sec:9}Summary and Conclusion}

In this work, we have investigated a certain class of time-delayed feedback systems with linear instantaneous and nonlinear delayed term for constant delay and modulated delay.
If the nonlinearity is chosen such that the resulting dynamical system possesses nonhyperbolic fixed-point solutions,
anomalous behavior such as anomalous diffusion, weak chaos, and weak ergodicty breaking can be found.
In the constant delay case, the observation of these phenomena is intricate because of the high dimension of the turbulent phases that reduces their probability of being trapped by the nonhyperbolic fixed points.
Once captured, the following slow escape from these fixed-point solutions can be described by center manifold theory and explains the emergence of residence time statistics with a diverging mean.
With the introduction of a slight modulation of the delay and the emergence of low-dimensional laminar chaos, this probability drastically increases
leading to an observation of the anomalous behavior already on short time scales.
For larger modulation amplitudes, residence time statistics is not just described by a single power law leading to a nontrivial scaling of the mean-squared displacement.
This more complex type of dynamical behavior could be reproduced by a two-dimensional iterated map.
The relation between the nontrivial residence time statistics and the MSD could be derived analytically with the help of a stochastic model.

The results reported in this paper are general in the sense that they only depend on the occurrence of the nonhyperbolic fixed-point solutions in this class of delay systems.
The responsible nonlinearities of the Pomeau-Manneville type are numerous and well known in literature \cite{geisel1984,zumofen1993,bel2006}.
A possible generalization could be achieved by ``transporting'' nonhyperbolic fixed-point solutions well known from the study of iterated maps \cite{geisel1985,zumofen1993,zumofen1995}
possibly leading to L\'evy walks in function space.

\begin{acknowledgments}
The authors gratefully acknowledge funding by the Deutsche Forschungsgemeinschaft (DFG, German Research Foundation) - 438881351;  456546951.
\end{acknowledgments}

\bibliography{references}

\begin{thebibliography}{90}%
\makeatletter
\providecommand \@ifxundefined [1]{%
 \@ifx{#1\undefined}
}%
\providecommand \@ifnum [1]{%
 \ifnum #1\expandafter \@firstoftwo
 \else \expandafter \@secondoftwo
 \fi
}%
\providecommand \@ifx [1]{%
 \ifx #1\expandafter \@firstoftwo
 \else \expandafter \@secondoftwo
 \fi
}%
\providecommand \natexlab [1]{#1}%
\providecommand \enquote  [1]{``#1''}%
\providecommand \bibnamefont  [1]{#1}%
\providecommand \bibfnamefont [1]{#1}%
\providecommand \citenamefont [1]{#1}%
\providecommand \href@noop [0]{\@secondoftwo}%
\providecommand \href [0]{\begingroup \@sanitize@url \@href}%
\providecommand \@href[1]{\@@startlink{#1}\@@href}%
\providecommand \@@href[1]{\endgroup#1\@@endlink}%
\providecommand \@sanitize@url [0]{\catcode `\\12\catcode `\$12\catcode
  `\&12\catcode `\#12\catcode `\^12\catcode `\_12\catcode `\%12\relax}%
\providecommand \@@startlink[1]{}%
\providecommand \@@endlink[0]{}%
\providecommand \url  [0]{\begingroup\@sanitize@url \@url }%
\providecommand \@url [1]{\endgroup\@href {#1}{\urlprefix }}%
\providecommand \urlprefix  [0]{URL }%
\providecommand \Eprint [0]{\href }%
\providecommand \doibase [0]{https://doi.org/}%
\providecommand \selectlanguage [0]{\@gobble}%
\providecommand \bibinfo  [0]{\@secondoftwo}%
\providecommand \bibfield  [0]{\@secondoftwo}%
\providecommand \translation [1]{[#1]}%
\providecommand \BibitemOpen [0]{}%
\providecommand \bibitemStop [0]{}%
\providecommand \bibitemNoStop [0]{.\EOS\space}%
\providecommand \EOS [0]{\spacefactor3000\relax}%
\providecommand \BibitemShut  [1]{\csname bibitem#1\endcsname}%
\let\auto@bib@innerbib\@empty
\bibitem [{\citenamefont {Geisel}\ and\ \citenamefont
  {Nierwetberg}(1982)}]{geisel1982}%
  \BibitemOpen
  \bibfield  {author} {\bibinfo {author} {\bibfnamefont {T.}~\bibnamefont
  {Geisel}}\ and\ \bibinfo {author} {\bibfnamefont {J.}~\bibnamefont
  {Nierwetberg}},\ }\bibfield  {title} {\bibinfo {title} {{Onset of Diffusion
  and Universal Scaling in Chaotic Systems}},\ }\href
  {https://doi.org/10.1103/PhysRevLett.48.7} {\bibfield  {journal} {\bibinfo
  {journal} {Phys. Rev. Lett.}\ }\textbf {\bibinfo {volume} {48}},\ \bibinfo
  {pages} {7} (\bibinfo {year} {1982})}\BibitemShut {NoStop}%
\bibitem [{\citenamefont {Schell}\ \emph {et~al.}(1982)\citenamefont {Schell},
  \citenamefont {Fraser},\ and\ \citenamefont {Kapral}}]{schell1982}%
  \BibitemOpen
  \bibfield  {author} {\bibinfo {author} {\bibfnamefont {M.}~\bibnamefont
  {Schell}}, \bibinfo {author} {\bibfnamefont {S.}~\bibnamefont {Fraser}},\
  and\ \bibinfo {author} {\bibfnamefont {R.}~\bibnamefont {Kapral}},\
  }\bibfield  {title} {\bibinfo {title} {{Diffusive dynamics in systems with
  translational symmetry: A one-dimensional-map model}},\ }\href
  {https://doi.org/10.1103/PhysRevA.26.504} {\bibfield  {journal} {\bibinfo
  {journal} {Phys. Rev. A}\ }\textbf {\bibinfo {volume} {26}},\ \bibinfo
  {pages} {504} (\bibinfo {year} {1982})}\BibitemShut {NoStop}%
\bibitem [{\citenamefont {Fujisaka}\ and\ \citenamefont
  {Grossmann}(1982)}]{fujisaka1982}%
  \BibitemOpen
  \bibfield  {author} {\bibinfo {author} {\bibfnamefont {H.}~\bibnamefont
  {Fujisaka}}\ and\ \bibinfo {author} {\bibfnamefont {S.}~\bibnamefont
  {Grossmann}},\ }\bibfield  {title} {\bibinfo {title} {{Chaos-Induced
  Diffusion in Nonlinear Discrete Dynamics}},\ }\href
  {https://doi.org/10.1007/BF01420589} {\bibfield  {journal} {\bibinfo
  {journal} {Z. Phys. B}\ }\textbf {\bibinfo {volume} {48}},\ \bibinfo {pages}
  {261} (\bibinfo {year} {1982})}\BibitemShut {NoStop}%
\bibitem [{\citenamefont {Bouchaud}\ and\ \citenamefont
  {Georges}(1990)}]{bouchaud1990}%
  \BibitemOpen
  \bibfield  {author} {\bibinfo {author} {\bibfnamefont {J.-P.}\ \bibnamefont
  {Bouchaud}}\ and\ \bibinfo {author} {\bibfnamefont {A.}~\bibnamefont
  {Georges}},\ }\bibfield  {title} {\bibinfo {title} {{Anomalous diffusion in
  disordered media: Statistical mechanisms, models and physical
  applications}},\ }\href {https://doi.org/10.1016/0370-1573(90)90099-N}
  {\bibfield  {journal} {\bibinfo  {journal} {Phys. Rep.}\ }\textbf {\bibinfo
  {volume} {195}},\ \bibinfo {pages} {127} (\bibinfo {year}
  {1990})}\BibitemShut {NoStop}%
\bibitem [{\citenamefont {Klages}\ \emph {et~al.}(2008)\citenamefont {Klages},
  \citenamefont {Radons},\ and\ \citenamefont {Sokolov}}]{klages2008}%
  \BibitemOpen
  \bibinfo {editor} {\bibfnamefont {R.}~\bibnamefont {Klages}}, \bibinfo
  {editor} {\bibfnamefont {G.}~\bibnamefont {Radons}},\ and\ \bibinfo {editor}
  {\bibfnamefont {I.~M.}\ \bibnamefont {Sokolov}},\ eds.,\ \href
  {https://doi.org/10.1002/9783527622979} {\emph {\bibinfo {title} {{Anomalous
  Transport: Foundations and Applications}}}},\ \bibinfo {edition} {1st}\ ed.\
  (\bibinfo  {publisher} {Wiley-VCH},\ \bibinfo {address} {Weinheim},\ \bibinfo
  {year} {2008})\BibitemShut {NoStop}%
\bibitem [{\citenamefont {Geisel}\ and\ \citenamefont
  {Thomae}(1984)}]{geisel1984}%
  \BibitemOpen
  \bibfield  {author} {\bibinfo {author} {\bibfnamefont {T.}~\bibnamefont
  {Geisel}}\ and\ \bibinfo {author} {\bibfnamefont {S.}~\bibnamefont
  {Thomae}},\ }\bibfield  {title} {\bibinfo {title} {{Anomalous Diffusion in
  Intermittent Chaotic Systems}},\ }\href
  {https://doi.org/10.1103/PhysRevLett.52.1936} {\bibfield  {journal} {\bibinfo
   {journal} {Phys. Rev. Lett.}\ }\textbf {\bibinfo {volume} {52}},\ \bibinfo
  {pages} {1936} (\bibinfo {year} {1984})}\BibitemShut {NoStop}%
\bibitem [{\citenamefont {Zumofen}\ and\ \citenamefont
  {Klafter}(1993)}]{zumofen1993}%
  \BibitemOpen
  \bibfield  {author} {\bibinfo {author} {\bibfnamefont {G.}~\bibnamefont
  {Zumofen}}\ and\ \bibinfo {author} {\bibfnamefont {J.}~\bibnamefont
  {Klafter}},\ }\bibfield  {title} {\bibinfo {title} {{Scale-invariant motion
  in intermittent chaotic systems}},\ }\href
  {https://doi.org/10.1103/PhysRevE.47.851} {\bibfield  {journal} {\bibinfo
  {journal} {Phys. Rev. E}\ }\textbf {\bibinfo {volume} {47}},\ \bibinfo
  {pages} {851} (\bibinfo {year} {1993})}\BibitemShut {NoStop}%
\bibitem [{\citenamefont {Bel}\ and\ \citenamefont {Barkai}(2006)}]{bel2006}%
  \BibitemOpen
  \bibfield  {author} {\bibinfo {author} {\bibfnamefont {G.}~\bibnamefont
  {Bel}}\ and\ \bibinfo {author} {\bibfnamefont {E.}~\bibnamefont {Barkai}},\
  }\bibfield  {title} {\bibinfo {title} {{Weak ergodicity breaking with
  deterministic dynamics}},\ }\href {https://doi.org/10.1209/epl/i2005-10501-8}
  {\bibfield  {journal} {\bibinfo  {journal} {Europhys. Lett.}\ }\textbf
  {\bibinfo {volume} {74}},\ \bibinfo {pages} {15} (\bibinfo {year}
  {2006})}\BibitemShut {NoStop}%
\bibitem [{\citenamefont {Chirikov}(1979)}]{chirikov1979}%
  \BibitemOpen
  \bibfield  {author} {\bibinfo {author} {\bibfnamefont {B.~V.}\ \bibnamefont
  {Chirikov}},\ }\bibfield  {title} {\bibinfo {title} {{A universal instability
  of many-dimensional oscillator systems}},\ }\href
  {https://doi.org/10.1016/0370-1573(79)90023-1} {\bibfield  {journal}
  {\bibinfo  {journal} {Phys. Rep.}\ }\textbf {\bibinfo {volume} {52}},\
  \bibinfo {pages} {263} (\bibinfo {year} {1979})}\BibitemShut {NoStop}%
\bibitem [{\citenamefont {Zacherl}\ \emph {et~al.}(1986)\citenamefont
  {Zacherl}, \citenamefont {Geisel}, \citenamefont {Nierwetberg},\ and\
  \citenamefont {Radons}}]{zacherl1986}%
  \BibitemOpen
  \bibfield  {author} {\bibinfo {author} {\bibfnamefont {A.}~\bibnamefont
  {Zacherl}}, \bibinfo {author} {\bibfnamefont {T.}~\bibnamefont {Geisel}},
  \bibinfo {author} {\bibfnamefont {J.}~\bibnamefont {Nierwetberg}},\ and\
  \bibinfo {author} {\bibfnamefont {G.}~\bibnamefont {Radons}},\ }\bibfield
  {title} {\bibinfo {title} {{Power spectra for anomalous diffusion in the
  extended Sinai billiard}},\ }\href
  {https://doi.org/10.1016/0375-9601(86)90568-2} {\bibfield  {journal}
  {\bibinfo  {journal} {Phys. Lett.}\ }\textbf {\bibinfo {volume} {114A}},\
  \bibinfo {pages} {317} (\bibinfo {year} {1986})}\BibitemShut {NoStop}%
\bibitem [{\citenamefont {Geisel}\ \emph {et~al.}(1987)\citenamefont {Geisel},
  \citenamefont {Zacherl},\ and\ \citenamefont {Radons}}]{geisel1987}%
  \BibitemOpen
  \bibfield  {author} {\bibinfo {author} {\bibfnamefont {T.}~\bibnamefont
  {Geisel}}, \bibinfo {author} {\bibfnamefont {A.}~\bibnamefont {Zacherl}},\
  and\ \bibinfo {author} {\bibfnamefont {G.}~\bibnamefont {Radons}},\
  }\bibfield  {title} {\bibinfo {title} {{Generic 1/f Noise in Chaotic
  Hamiltonian Dynamics}},\ }\href {https://doi.org/10.1103/PhysRevLett.59.2503}
  {\bibfield  {journal} {\bibinfo  {journal} {Phys. Rev. Lett.}\ }\textbf
  {\bibinfo {volume} {59}},\ \bibinfo {pages} {2503} (\bibinfo {year}
  {1987})}\BibitemShut {NoStop}%
\bibitem [{\citenamefont {Lichtenberg}\ and\ \citenamefont
  {Lieberman}(1992)}]{lichtenberg1992}%
  \BibitemOpen
  \bibfield  {author} {\bibinfo {author} {\bibfnamefont {A.~J.}\ \bibnamefont
  {Lichtenberg}}\ and\ \bibinfo {author} {\bibfnamefont {M.~A.}\ \bibnamefont
  {Lieberman}},\ }\href {https://doi.org/10.1007/978-1-4757-2184-3} {\emph
  {\bibinfo {title} {{Regular and Chaotic Dynamics}}}},\ \bibinfo {edition}
  {2nd}\ ed.\ (\bibinfo  {publisher} {Springer},\ \bibinfo {address} {New
  York},\ \bibinfo {year} {1992})\BibitemShut {NoStop}%
\bibitem [{\citenamefont {Zumofen}\ and\ \citenamefont
  {Klafter}(1994)}]{zumofen1994}%
  \BibitemOpen
  \bibfield  {author} {\bibinfo {author} {\bibfnamefont {G.}~\bibnamefont
  {Zumofen}}\ and\ \bibinfo {author} {\bibfnamefont {J.}~\bibnamefont
  {Klafter}},\ }\bibfield  {title} {\bibinfo {title} {{Random Walks in the
  Standard Map}},\ }\href {https://doi.org/10.1209/0295-5075/25/8/002}
  {\bibfield  {journal} {\bibinfo  {journal} {Europhys. Lett.}\ }\textbf
  {\bibinfo {volume} {25}},\ \bibinfo {pages} {565} (\bibinfo {year}
  {1994})}\BibitemShut {NoStop}%
\bibitem [{\citenamefont {Hale}\ and\ \citenamefont {{Verduyn
  Lunel}}(1993)}]{hale1993}%
  \BibitemOpen
  \bibfield  {author} {\bibinfo {author} {\bibfnamefont {J.~K.}\ \bibnamefont
  {Hale}}\ and\ \bibinfo {author} {\bibfnamefont {S.~M.}\ \bibnamefont
  {{Verduyn Lunel}}},\ }\href {https://doi.org/10.1007/978-1-4612-4342-7}
  {\emph {\bibinfo {title} {{Introduction to Functional Differential
  Equations}}}},\ \bibinfo {edition} {1st}\ ed.\ (\bibinfo  {publisher}
  {Springer},\ \bibinfo {address} {New York},\ \bibinfo {year}
  {1993})\BibitemShut {NoStop}%
\bibitem [{\citenamefont {Diekmann}\ \emph {et~al.}(1995)\citenamefont
  {Diekmann}, \citenamefont {{van Gils}}, \citenamefont {{Verduyn Lunel}},\
  and\ \citenamefont {Walther}}]{diekmann1995}%
  \BibitemOpen
  \bibfield  {author} {\bibinfo {author} {\bibfnamefont {O.}~\bibnamefont
  {Diekmann}}, \bibinfo {author} {\bibfnamefont {S.~A.}\ \bibnamefont {{van
  Gils}}}, \bibinfo {author} {\bibfnamefont {S.~M.}\ \bibnamefont {{Verduyn
  Lunel}}},\ and\ \bibinfo {author} {\bibfnamefont {H.-O.}\ \bibnamefont
  {Walther}},\ }\href {https://doi.org/10.1007/978-1-4612-4206-2} {\emph
  {\bibinfo {title} {{Delay Equations: Functional-, Complex-, and Nonlinear
  Analysis}}}},\ \bibinfo {edition} {1st}\ ed.\ (\bibinfo  {publisher}
  {Springer},\ \bibinfo {address} {New York},\ \bibinfo {year}
  {1995})\BibitemShut {NoStop}%
\bibitem [{\citenamefont {Driver}(1963)}]{driver1963}%
  \BibitemOpen
  \bibfield  {author} {\bibinfo {author} {\bibfnamefont {R.~D.}\ \bibnamefont
  {Driver}},\ }\bibfield  {title} {\bibinfo {title} {{A two-body problem of
  classical electrodynamics: the one-dimensional case}},\ }\href
  {https://doi.org/10.1016/0003-4916(63)90227-6} {\bibfield  {journal}
  {\bibinfo  {journal} {Ann. Phys. (NY)}\ }\textbf {\bibinfo {volume} {21}},\
  \bibinfo {pages} {122} (\bibinfo {year} {1963})}\BibitemShut {NoStop}%
\bibitem [{\citenamefont {L\'opez}(2020)}]{lopez2020}%
  \BibitemOpen
  \bibfield  {author} {\bibinfo {author} {\bibfnamefont {A.~G.}\ \bibnamefont
  {L\'opez}},\ }\bibfield  {title} {\bibinfo {title} {{On an electrodynamic
  origin of quantum fluctuations}},\ }\href
  {https://doi.org/10.1007/s11071-020-05928-5} {\bibfield  {journal} {\bibinfo
  {journal} {Nonlinear Dyn.}\ }\textbf {\bibinfo {volume} {102}},\ \bibinfo
  {pages} {621} (\bibinfo {year} {2020})}\BibitemShut {NoStop}%
\bibitem [{\citenamefont {Gopalsamy}(1992)}]{gopalsamy1992}%
  \BibitemOpen
  \bibfield  {author} {\bibinfo {author} {\bibfnamefont {K.}~\bibnamefont
  {Gopalsamy}},\ }\href {https://doi.org/10.1007/978-94-015-7920-9} {\emph
  {\bibinfo {title} {{Stability and Oscillations in Delay Differential
  Equations of Population Dynamics}}}},\ \bibinfo {edition} {1st}\ ed.\
  (\bibinfo  {publisher} {Springer},\ \bibinfo {address} {Dordrecht},\ \bibinfo
  {year} {1992})\BibitemShut {NoStop}%
\bibitem [{\citenamefont {Kuang}(1993)}]{kuang1993}%
  \BibitemOpen
  \bibfield  {author} {\bibinfo {author} {\bibfnamefont {Y.}~\bibnamefont
  {Kuang}},\ }\href@noop {} {\emph {\bibinfo {title} {{Delay Differential
  Equations With Applications in Population Dynamics}}}},\ \bibinfo {edition}
  {1st}\ ed.\ (\bibinfo  {publisher} {Academic Press},\ \bibinfo {address} {San
  Diego},\ \bibinfo {year} {1993})\BibitemShut {NoStop}%
\bibitem [{\citenamefont {Salpeter}\ and\ \citenamefont
  {Salpeter}(1998)}]{salpeter1998}%
  \BibitemOpen
  \bibfield  {author} {\bibinfo {author} {\bibfnamefont {E.~E.}\ \bibnamefont
  {Salpeter}}\ and\ \bibinfo {author} {\bibfnamefont {S.~R.}\ \bibnamefont
  {Salpeter}},\ }\bibfield  {title} {\bibinfo {title} {{Mathematical Model for
  the Epidemiology of Tuberculosis, with Estimates of the Reproductive Number
  and Infection-Delay Function}},\ }\href
  {https://doi.org/10.1093/oxfordjournals.aje.a009463} {\bibfield  {journal}
  {\bibinfo  {journal} {Am. J. Epidemiol.}\ }\textbf {\bibinfo {volume}
  {142}},\ \bibinfo {pages} {398} (\bibinfo {year} {1998})}\BibitemShut
  {NoStop}%
\bibitem [{\citenamefont {Kajiwara}\ \emph {et~al.}(2012)\citenamefont
  {Kajiwara}, \citenamefont {Sasaki},\ and\ \citenamefont
  {Takeuchi}}]{kajiwara2012}%
  \BibitemOpen
  \bibfield  {author} {\bibinfo {author} {\bibfnamefont {T.}~\bibnamefont
  {Kajiwara}}, \bibinfo {author} {\bibfnamefont {T.}~\bibnamefont {Sasaki}},\
  and\ \bibinfo {author} {\bibfnamefont {Y.}~\bibnamefont {Takeuchi}},\
  }\bibfield  {title} {\bibinfo {title} {{Construction of Lyapunov functionals
  for delay differential equations in virology and epidemiology}},\ }\href
  {https://doi.org/10.1016/j.nonrwa.2011.12.011} {\bibfield  {journal}
  {\bibinfo  {journal} {Nonl. Anal. RWA}\ }\textbf {\bibinfo {volume} {13}},\
  \bibinfo {pages} {1802} (\bibinfo {year} {2012})}\BibitemShut {NoStop}%
\bibitem [{\citenamefont {Ikeda}\ \emph {et~al.}(1980)\citenamefont {Ikeda},
  \citenamefont {Daido},\ and\ \citenamefont {Akimoto}}]{ikeda1980}%
  \BibitemOpen
  \bibfield  {author} {\bibinfo {author} {\bibfnamefont {K.}~\bibnamefont
  {Ikeda}}, \bibinfo {author} {\bibfnamefont {H.}~\bibnamefont {Daido}},\ and\
  \bibinfo {author} {\bibfnamefont {O.}~\bibnamefont {Akimoto}},\ }\bibfield
  {title} {\bibinfo {title} {{Optical Turbulence: Chaotic Behavior of
  Transmitted Light from a Ring Cavity}},\ }\href
  {https://doi.org/10.1103/PhysRevLett.45.709} {\bibfield  {journal} {\bibinfo
  {journal} {Phys. Rev. Lett.}\ }\textbf {\bibinfo {volume} {45}},\ \bibinfo
  {pages} {709} (\bibinfo {year} {1980})}\BibitemShut {NoStop}%
\bibitem [{\citenamefont {Erneux}(2009)}]{erneux2009}%
  \BibitemOpen
  \bibfield  {author} {\bibinfo {author} {\bibfnamefont {T.}~\bibnamefont
  {Erneux}},\ }\href {https://doi.org/10.1007/978-0-387-74372-1} {\emph
  {\bibinfo {title} {{Applied Delay Differential Equations}}}},\ \bibinfo
  {edition} {1st}\ ed.\ (\bibinfo  {publisher} {Springer},\ \bibinfo {address}
  {New York},\ \bibinfo {year} {2009})\BibitemShut {NoStop}%
\bibitem [{\citenamefont {Insperger}\ and\ \citenamefont
  {St\'ep\'an}(2011)}]{insperger2011}%
  \BibitemOpen
  \bibfield  {author} {\bibinfo {author} {\bibfnamefont {T.}~\bibnamefont
  {Insperger}}\ and\ \bibinfo {author} {\bibfnamefont {G.}~\bibnamefont
  {St\'ep\'an}},\ }\href {https://doi.org/10.1007/978-1-4614-0335-7} {\emph
  {\bibinfo {title} {{Semi-Discretization for Time-Delay Systems: Stability and
  Engineering Applications}}}},\ \bibinfo {edition} {1st}\ ed.\ (\bibinfo
  {publisher} {Springer},\ \bibinfo {address} {New York},\ \bibinfo {year}
  {2011})\BibitemShut {NoStop}%
\bibitem [{\citenamefont {Wischert}\ \emph {et~al.}(1994)\citenamefont
  {Wischert}, \citenamefont {Wunderlin}, \citenamefont {Pelster}, \citenamefont
  {Olivier},\ and\ \citenamefont {Groslambert}}]{wischert1994}%
  \BibitemOpen
  \bibfield  {author} {\bibinfo {author} {\bibfnamefont {W.}~\bibnamefont
  {Wischert}}, \bibinfo {author} {\bibfnamefont {A.}~\bibnamefont {Wunderlin}},
  \bibinfo {author} {\bibfnamefont {A.}~\bibnamefont {Pelster}}, \bibinfo
  {author} {\bibfnamefont {M.}~\bibnamefont {Olivier}},\ and\ \bibinfo {author}
  {\bibfnamefont {J.}~\bibnamefont {Groslambert}},\ }\bibfield  {title}
  {\bibinfo {title} {{Delay-induced instabilities in nonlinear feedback
  systems}},\ }\href {https://doi.org/10.1103/PhysRevE.49.203} {\bibfield
  {journal} {\bibinfo  {journal} {Phys. Rev. E}\ }\textbf {\bibinfo {volume}
  {49}},\ \bibinfo {pages} {203} (\bibinfo {year} {1994})}\BibitemShut
  {NoStop}%
\bibitem [{\citenamefont {Schanz}\ and\ \citenamefont
  {Pelster}(2003)}]{schanz2003}%
  \BibitemOpen
  \bibfield  {author} {\bibinfo {author} {\bibfnamefont {M.}~\bibnamefont
  {Schanz}}\ and\ \bibinfo {author} {\bibfnamefont {A.}~\bibnamefont
  {Pelster}},\ }\bibfield  {title} {\bibinfo {title} {{Analytical and numerical
  investigations of the phase-locked loop with time delay}},\ }\href
  {https://doi.org/10.1103/PhysRevE.67.056205} {\bibfield  {journal} {\bibinfo
  {journal} {Phys. Rev. E}\ }\textbf {\bibinfo {volume} {67}},\ \bibinfo
  {pages} {056205} (\bibinfo {year} {2003})}\BibitemShut {NoStop}%
\bibitem [{\citenamefont {Sprott}(2007)}]{sprott2007}%
  \BibitemOpen
  \bibfield  {author} {\bibinfo {author} {\bibfnamefont {J.~C.}\ \bibnamefont
  {Sprott}},\ }\bibfield  {title} {\bibinfo {title} {{A simple chaotic delay
  differential equation}},\ }\href
  {https://doi.org/10.1016/j.physleta.2007.01.083} {\bibfield  {journal}
  {\bibinfo  {journal} {Phys. Lett. A}\ }\textbf {\bibinfo {volume} {366}},\
  \bibinfo {pages} {397} (\bibinfo {year} {2007})}\BibitemShut {NoStop}%
\bibitem [{\citenamefont {Lei}\ and\ \citenamefont {Mackey}(2011)}]{lei2011}%
  \BibitemOpen
  \bibfield  {author} {\bibinfo {author} {\bibfnamefont {J.}~\bibnamefont
  {Lei}}\ and\ \bibinfo {author} {\bibfnamefont {M.~C.}\ \bibnamefont
  {Mackey}},\ }\bibfield  {title} {\bibinfo {title} {{Deterministic Brownian
  motion generated from differential delay equations}},\ }\href
  {https://doi.org/10.1103/PhysRevE.84.041105} {\bibfield  {journal} {\bibinfo
  {journal} {Phys. Rev. E}\ }\textbf {\bibinfo {volume} {84}},\ \bibinfo
  {pages} {041105} (\bibinfo {year} {2011})}\BibitemShut {NoStop}%
\bibitem [{\citenamefont {Dao}(2013)}]{dao2013_1}%
  \BibitemOpen
  \bibfield  {author} {\bibinfo {author} {\bibfnamefont {H.~T.~L.}\
  \bibnamefont {Dao}},\ }\emph {\bibinfo {title} {{Complex dynamics of a
  microwave time-delayed feedback loop}}},\ \href
  {https://doi.org/10.13016/M21P4C} {\bibinfo {type} {Dissertation}},\ \bibinfo
   {school} {Graduate School of the University of Maryland}, \bibinfo {address}
  {College Park, Maryland} (\bibinfo {year} {2013})\BibitemShut {NoStop}%
\bibitem [{\citenamefont {Dao}\ \emph {et~al.}(2013)\citenamefont {Dao},
  \citenamefont {Rodgers},\ and\ \citenamefont {Murphy}}]{dao2013_2}%
  \BibitemOpen
  \bibfield  {author} {\bibinfo {author} {\bibfnamefont {H.}~\bibnamefont
  {Dao}}, \bibinfo {author} {\bibfnamefont {J.~C.}\ \bibnamefont {Rodgers}},\
  and\ \bibinfo {author} {\bibfnamefont {T.~E.}\ \bibnamefont {Murphy}},\
  }\bibfield  {title} {\bibinfo {title} {{Chaotic dynamics of a
  frequency-modulated microwave oscillator with time-delayed feedback}},\
  }\href {https://doi.org/10.1063/1.4772970} {\bibfield  {journal} {\bibinfo
  {journal} {Chaos}\ }\textbf {\bibinfo {volume} {23}},\ \bibinfo {pages}
  {013101} (\bibinfo {year} {2013})}\BibitemShut {NoStop}%
\bibitem [{\citenamefont {Mackey}\ and\ \citenamefont
  {Tyran-Kami\'nska}(2021)}]{mackey2021}%
  \BibitemOpen
  \bibfield  {author} {\bibinfo {author} {\bibfnamefont {M.~C.}\ \bibnamefont
  {Mackey}}\ and\ \bibinfo {author} {\bibfnamefont {M.}~\bibnamefont
  {Tyran-Kami\'nska}},\ }\bibfield  {title} {\bibinfo {title} {{How can we
  describe density evolution under delayed dynamics?}},\ }\href
  {https://doi.org/10.1063/5.0038310} {\bibfield  {journal} {\bibinfo
  {journal} {Chaos}\ }\textbf {\bibinfo {volume} {31}},\ \bibinfo {pages}
  {043114} (\bibinfo {year} {2021})}\BibitemShut {NoStop}%
\bibitem [{\citenamefont {Albers}\ \emph
  {et~al.}(2019{\natexlab{a}})\citenamefont {Albers}, \citenamefont
  {Cisternas},\ and\ \citenamefont {Radons}}]{albers2019_1}%
  \BibitemOpen
  \bibfield  {author} {\bibinfo {author} {\bibfnamefont {T.}~\bibnamefont
  {Albers}}, \bibinfo {author} {\bibfnamefont {J.}~\bibnamefont {Cisternas}},\
  and\ \bibinfo {author} {\bibfnamefont {G.}~\bibnamefont {Radons}},\
  }\bibfield  {title} {\bibinfo {title} {{A new kind of chaotic diffusion:
  anti-persistent random walks of explosive dissipative solitons}},\ }\href
  {https://doi.org/10.1088/1367-2630/ab4884} {\bibfield  {journal} {\bibinfo
  {journal} {New J. Phys.}\ }\textbf {\bibinfo {volume} {21}},\ \bibinfo
  {pages} {103034} (\bibinfo {year} {2019}{\natexlab{a}})}\BibitemShut
  {NoStop}%
\bibitem [{\citenamefont {Albers}\ \emph
  {et~al.}(2019{\natexlab{b}})\citenamefont {Albers}, \citenamefont
  {Cisternas},\ and\ \citenamefont {Radons}}]{albers2019_2}%
  \BibitemOpen
  \bibfield  {author} {\bibinfo {author} {\bibfnamefont {T.}~\bibnamefont
  {Albers}}, \bibinfo {author} {\bibfnamefont {J.}~\bibnamefont {Cisternas}},\
  and\ \bibinfo {author} {\bibfnamefont {G.}~\bibnamefont {Radons}},\
  }\bibfield  {title} {\bibinfo {title} {{A hidden Markov model for the
  dynamics of diffusing dissipative solitons}},\ }\href
  {https://doi.org/10.1088/1742-5468/ab3986} {\bibfield  {journal} {\bibinfo
  {journal} {J. Stat. Mech.}\ }\textbf {\bibinfo {volume} {2019}},\ \bibinfo
  {pages} {094013} (\bibinfo {year} {2019}{\natexlab{b}})}\BibitemShut
  {NoStop}%
\bibitem [{\citenamefont {Albers}\ \emph
  {et~al.}(2022{\natexlab{a}})\citenamefont {Albers}, \citenamefont
  {M\"uller-Bender}, \citenamefont {Hille},\ and\ \citenamefont
  {Radons}}]{albers2022_1}%
  \BibitemOpen
  \bibfield  {author} {\bibinfo {author} {\bibfnamefont {T.}~\bibnamefont
  {Albers}}, \bibinfo {author} {\bibfnamefont {D.}~\bibnamefont
  {M\"uller-Bender}}, \bibinfo {author} {\bibfnamefont {L.}~\bibnamefont
  {Hille}},\ and\ \bibinfo {author} {\bibfnamefont {G.}~\bibnamefont
  {Radons}},\ }\bibfield  {title} {\bibinfo {title} {{Chaotic Diffusion in
  Delay Systems: Giant Enhancement by Time Lag Modulation}},\ }\href
  {https://doi.org/10.1103/PhysRevLett.128.074101} {\bibfield  {journal}
  {\bibinfo  {journal} {Phys. Rev. Lett.}\ }\textbf {\bibinfo {volume} {128}},\
  \bibinfo {pages} {074101} (\bibinfo {year} {2022}{\natexlab{a}})}\BibitemShut
  {NoStop}%
\bibitem [{\citenamefont {Albers}\ \emph
  {et~al.}(2022{\natexlab{b}})\citenamefont {Albers}, \citenamefont
  {M\"uller-Bender},\ and\ \citenamefont {Radons}}]{albers2022_2}%
  \BibitemOpen
  \bibfield  {author} {\bibinfo {author} {\bibfnamefont {T.}~\bibnamefont
  {Albers}}, \bibinfo {author} {\bibfnamefont {D.}~\bibnamefont
  {M\"uller-Bender}},\ and\ \bibinfo {author} {\bibfnamefont {G.}~\bibnamefont
  {Radons}},\ }\bibfield  {title} {\bibinfo {title} {{Antipersistent random
  walks in time-delayed systems}},\ }\href
  {https://doi.org/10.1103/PhysRevE.105.064212} {\bibfield  {journal} {\bibinfo
   {journal} {Phys. Rev. E}\ }\textbf {\bibinfo {volume} {105}},\ \bibinfo
  {pages} {064212} (\bibinfo {year} {2022}{\natexlab{b}})}\BibitemShut
  {NoStop}%
\bibitem [{\citenamefont {Cisternas}\ \emph {et~al.}(2016)\citenamefont
  {Cisternas}, \citenamefont {Descalzi}, \citenamefont {Albers},\ and\
  \citenamefont {Radons}}]{cisternas2016}%
  \BibitemOpen
  \bibfield  {author} {\bibinfo {author} {\bibfnamefont {J.}~\bibnamefont
  {Cisternas}}, \bibinfo {author} {\bibfnamefont {O.}~\bibnamefont {Descalzi}},
  \bibinfo {author} {\bibfnamefont {T.}~\bibnamefont {Albers}},\ and\ \bibinfo
  {author} {\bibfnamefont {G.}~\bibnamefont {Radons}},\ }\bibfield  {title}
  {\bibinfo {title} {{Anomalous Diffusion of Dissipative Solitons in the
  Cubic-Quintic Complex Ginzburg-Landau Equation in Two Spatial Dimensions}},\
  }\href {https://doi.org/10.1103/PhysRevLett.116.203901} {\bibfield  {journal}
  {\bibinfo  {journal} {Phys. Rev. Lett.}\ }\textbf {\bibinfo {volume} {116}},\
  \bibinfo {pages} {203901} (\bibinfo {year} {2016})}\BibitemShut {NoStop}%
\bibitem [{\citenamefont {Cisternas}\ \emph {et~al.}(2018)\citenamefont
  {Cisternas}, \citenamefont {Albers},\ and\ \citenamefont
  {Radons}}]{cisternas2018}%
  \BibitemOpen
  \bibfield  {author} {\bibinfo {author} {\bibfnamefont {J.}~\bibnamefont
  {Cisternas}}, \bibinfo {author} {\bibfnamefont {T.}~\bibnamefont {Albers}},\
  and\ \bibinfo {author} {\bibfnamefont {G.}~\bibnamefont {Radons}},\
  }\bibfield  {title} {\bibinfo {title} {{Normal and anomalous random walks of
  2-d solitons}},\ }\href {https://doi.org/10.1063/1.5021586} {\bibfield
  {journal} {\bibinfo  {journal} {Chaos}\ }\textbf {\bibinfo {volume} {28}},\
  \bibinfo {pages} {075505} (\bibinfo {year} {2018})}\BibitemShut {NoStop}%
\bibitem [{\citenamefont {Scher}\ and\ \citenamefont
  {Montroll}(1975)}]{scher1975}%
  \BibitemOpen
  \bibfield  {author} {\bibinfo {author} {\bibfnamefont {H.}~\bibnamefont
  {Scher}}\ and\ \bibinfo {author} {\bibfnamefont {E.~W.}\ \bibnamefont
  {Montroll}},\ }\bibfield  {title} {\bibinfo {title} {{Anomalous transit-time
  dispersion in amorphous solids}},\ }\href
  {https://doi.org/10.1103/PhysRevB.12.2455} {\bibfield  {journal} {\bibinfo
  {journal} {Phys. Rev. B}\ }\textbf {\bibinfo {volume} {12}},\ \bibinfo
  {pages} {2455} (\bibinfo {year} {1975})}\BibitemShut {NoStop}%
\bibitem [{\citenamefont {Sch\"utz}\ \emph {et~al.}(1997)\citenamefont
  {Sch\"utz}, \citenamefont {Schindler},\ and\ \citenamefont
  {Schmidt}}]{schuetz1997}%
  \BibitemOpen
  \bibfield  {author} {\bibinfo {author} {\bibfnamefont {G.~J.}\ \bibnamefont
  {Sch\"utz}}, \bibinfo {author} {\bibfnamefont {H.}~\bibnamefont
  {Schindler}},\ and\ \bibinfo {author} {\bibfnamefont {T.}~\bibnamefont
  {Schmidt}},\ }\bibfield  {title} {\bibinfo {title} {{Single-Molecule
  Microscopy on Model Membranes Reveals Anomalous Diffusion}},\ }\href
  {https://doi.org/10.1016/S0006-3495(97)78139-6} {\bibfield  {journal}
  {\bibinfo  {journal} {Biophys. J.}\ }\textbf {\bibinfo {volume} {73}},\
  \bibinfo {pages} {1073} (\bibinfo {year} {1997})}\BibitemShut {NoStop}%
\bibitem [{\citenamefont {Krapf}(2015)}]{krapf2015}%
  \BibitemOpen
  \bibfield  {author} {\bibinfo {author} {\bibfnamefont {D.}~\bibnamefont
  {Krapf}},\ }\bibfield  {title} {\bibinfo {title} {{Mechanisms Underlying
  Anomalous Diffusion in the Plasma Membrane}},\ }in\ \href
  {https://doi.org/10.1016/bs.ctm.2015.03.002} {\emph {\bibinfo {booktitle}
  {Lipid Domains}}},\ \bibinfo {series} {Current Topics in Membranes},
  Vol.~\bibinfo {volume} {75},\ \bibinfo {editor} {edited by\ \bibinfo {editor}
  {\bibfnamefont {A.~K.}\ \bibnamefont {Kenworthy}}}\ (\bibinfo  {publisher}
  {Elsevier},\ \bibinfo {year} {2015})\ Chap.~\bibinfo {chapter} {5}, p.\
  \bibinfo {pages} {167}\BibitemShut {NoStop}%
\bibitem [{\citenamefont {Gefen}\ \emph {et~al.}(1983)\citenamefont {Gefen},
  \citenamefont {Aharony},\ and\ \citenamefont {Alexander}}]{gefen1983}%
  \BibitemOpen
  \bibfield  {author} {\bibinfo {author} {\bibfnamefont {Y.}~\bibnamefont
  {Gefen}}, \bibinfo {author} {\bibfnamefont {A.}~\bibnamefont {Aharony}},\
  and\ \bibinfo {author} {\bibfnamefont {S.}~\bibnamefont {Alexander}},\
  }\bibfield  {title} {\bibinfo {title} {{Anomalous Diffusion on Percolating
  Clusters}},\ }\href {https://doi.org/10.1103/PhysRevLett.50.77} {\bibfield
  {journal} {\bibinfo  {journal} {Phys. Rev. Lett.}\ }\textbf {\bibinfo
  {volume} {50}},\ \bibinfo {pages} {77} (\bibinfo {year} {1983})}\BibitemShut
  {NoStop}%
\bibitem [{\citenamefont {Havlin}\ and\ \citenamefont
  {Ben-Avraham}(1987)}]{havlin1987}%
  \BibitemOpen
  \bibfield  {author} {\bibinfo {author} {\bibfnamefont {S.}~\bibnamefont
  {Havlin}}\ and\ \bibinfo {author} {\bibfnamefont {D.}~\bibnamefont
  {Ben-Avraham}},\ }\bibfield  {title} {\bibinfo {title} {{Diffusion in
  disordered media}},\ }\href {https://doi.org/10.1080/00018738700101072}
  {\bibfield  {journal} {\bibinfo  {journal} {Adv. Phys.}\ }\textbf {\bibinfo
  {volume} {36}},\ \bibinfo {pages} {695} (\bibinfo {year} {1987})}\BibitemShut
  {NoStop}%
\bibitem [{\citenamefont {Montroll}\ and\ \citenamefont
  {Weiss}(1965)}]{montroll1965}%
  \BibitemOpen
  \bibfield  {author} {\bibinfo {author} {\bibfnamefont {E.~W.}\ \bibnamefont
  {Montroll}}\ and\ \bibinfo {author} {\bibfnamefont {G.~H.}\ \bibnamefont
  {Weiss}},\ }\bibfield  {title} {\bibinfo {title} {{Random Walks on
  Lattices}},\ }\href {https://doi.org/10.1063/1.1704269} {\bibfield  {journal}
  {\bibinfo  {journal} {J. Math. Phys. (N.Y.)}\ }\textbf {\bibinfo {volume}
  {6}},\ \bibinfo {pages} {167} (\bibinfo {year} {1965})}\BibitemShut {NoStop}%
\bibitem [{\citenamefont {Metzler}\ and\ \citenamefont
  {Klafter}(2000)}]{metzler2000}%
  \BibitemOpen
  \bibfield  {author} {\bibinfo {author} {\bibfnamefont {R.}~\bibnamefont
  {Metzler}}\ and\ \bibinfo {author} {\bibfnamefont {J.}~\bibnamefont
  {Klafter}},\ }\bibfield  {title} {\bibinfo {title} {{The random walk's guide
  to anomalous diffusion: a fractional dynamics approach}},\ }\href
  {https://doi.org/10.1016/S0370-1573(00)00070-3} {\bibfield  {journal}
  {\bibinfo  {journal} {Phys. Rep.}\ }\textbf {\bibinfo {volume} {339}},\
  \bibinfo {pages} {1} (\bibinfo {year} {2000})}\BibitemShut {NoStop}%
\bibitem [{\citenamefont {Mandelbrot}\ and\ \citenamefont
  {Ness}(1968)}]{mandelbrot1968}%
  \BibitemOpen
  \bibfield  {author} {\bibinfo {author} {\bibfnamefont {B.~B.}\ \bibnamefont
  {Mandelbrot}}\ and\ \bibinfo {author} {\bibfnamefont {J.~W.~V.}\ \bibnamefont
  {Ness}},\ }\bibfield  {title} {\bibinfo {title} {{Fractional Brownian
  Motions, Fractional Noises and Applications}},\ }\href
  {https://doi.org/10.1137/1010093} {\bibfield  {journal} {\bibinfo  {journal}
  {SIAM Rev.}\ }\textbf {\bibinfo {volume} {10}},\ \bibinfo {pages} {422}
  (\bibinfo {year} {1968})}\BibitemShut {NoStop}%
\bibitem [{\citenamefont {Albers}\ \emph {et~al.}(2025)\citenamefont {Albers},
  \citenamefont {Hille}, \citenamefont {M\"uller-Bender},\ and\ \citenamefont
  {Radons}}]{albers2025}%
  \BibitemOpen
  \bibfield  {author} {\bibinfo {author} {\bibfnamefont {T.}~\bibnamefont
  {Albers}}, \bibinfo {author} {\bibfnamefont {L.}~\bibnamefont {Hille}},
  \bibinfo {author} {\bibfnamefont {D.}~\bibnamefont {M\"uller-Bender}},\ and\
  \bibinfo {author} {\bibfnamefont {G.}~\bibnamefont {Radons}},\ }\bibfield
  {title} {\bibinfo {title} {{Weak chaos, anomalous diffusion, and weak
  ergodicity breaking in systems with delay}},\ }\href
  {https://doi.org/10.1103/qqcs-7pll} {\bibfield  {journal} {\bibinfo
  {journal} {Phys. Rev. E}\ }\textbf {\bibinfo {volume} {112}},\ \bibinfo
  {pages} {L042201} (\bibinfo {year} {2025})}\BibitemShut {NoStop}%
\bibitem [{\citenamefont {Korabel}\ and\ \citenamefont
  {Barkai}(2009)}]{korabel2009}%
  \BibitemOpen
  \bibfield  {author} {\bibinfo {author} {\bibfnamefont {N.}~\bibnamefont
  {Korabel}}\ and\ \bibinfo {author} {\bibfnamefont {E.}~\bibnamefont
  {Barkai}},\ }\bibfield  {title} {\bibinfo {title} {{Pesin-Type Identity for
  Intermittent Dynamics with a Zero Lyaponov Exponent}},\ }\href
  {https://doi.org/10.1103/PhysRevLett.102.050601} {\bibfield  {journal}
  {\bibinfo  {journal} {Phys. Rev. Lett.}\ }\textbf {\bibinfo {volume} {102}},\
  \bibinfo {pages} {050601} (\bibinfo {year} {2009})}\BibitemShut {NoStop}%
\bibitem [{\citenamefont {Korabel}\ and\ \citenamefont
  {Barkai}(2010)}]{korabel2010}%
  \BibitemOpen
  \bibfield  {author} {\bibinfo {author} {\bibfnamefont {N.}~\bibnamefont
  {Korabel}}\ and\ \bibinfo {author} {\bibfnamefont {E.}~\bibnamefont
  {Barkai}},\ }\bibfield  {title} {\bibinfo {title} {{Separation of
  trajectories and its relation to entropy for intermittent systems with a zero
  Lyapunov exponent}},\ }\href {https://doi.org/10.1103/PhysRevE.82.016209}
  {\bibfield  {journal} {\bibinfo  {journal} {Phys. Rev. E}\ }\textbf {\bibinfo
  {volume} {82}},\ \bibinfo {pages} {016209} (\bibinfo {year}
  {2010})}\BibitemShut {NoStop}%
\bibitem [{\citenamefont {Bouchaud}(1992)}]{bouchaud1992}%
  \BibitemOpen
  \bibfield  {author} {\bibinfo {author} {\bibfnamefont {J.~P.}\ \bibnamefont
  {Bouchaud}},\ }\bibfield  {title} {\bibinfo {title} {{Weak ergodicity
  breaking and aging in disordered systems}},\ }\href
  {https://doi.org/10.1051/jp1:1992238} {\bibfield  {journal} {\bibinfo
  {journal} {J. Phys. I (France)}\ }\textbf {\bibinfo {volume} {2}},\ \bibinfo
  {pages} {1705} (\bibinfo {year} {1992})}\BibitemShut {NoStop}%
\bibitem [{\citenamefont {Lubelski}\ \emph {et~al.}(2008)\citenamefont
  {Lubelski}, \citenamefont {Sokolov},\ and\ \citenamefont
  {Klafter}}]{lubelski2008}%
  \BibitemOpen
  \bibfield  {author} {\bibinfo {author} {\bibfnamefont {A.}~\bibnamefont
  {Lubelski}}, \bibinfo {author} {\bibfnamefont {I.~M.}\ \bibnamefont
  {Sokolov}},\ and\ \bibinfo {author} {\bibfnamefont {J.}~\bibnamefont
  {Klafter}},\ }\bibfield  {title} {\bibinfo {title} {{Nonergodicity Mimics
  Inhomogeneity in Single Particle Tracking}},\ }\href
  {https://doi.org/10.1103/PhysRevLett.100.250602} {\bibfield  {journal}
  {\bibinfo  {journal} {Phys. Rev. Lett.}\ }\textbf {\bibinfo {volume} {100}},\
  \bibinfo {pages} {250602} (\bibinfo {year} {2008})}\BibitemShut {NoStop}%
\bibitem [{\citenamefont {He}\ \emph {et~al.}(2008)\citenamefont {He},
  \citenamefont {Burov}, \citenamefont {Metzler},\ and\ \citenamefont
  {Barkai}}]{he2008}%
  \BibitemOpen
  \bibfield  {author} {\bibinfo {author} {\bibfnamefont {Y.}~\bibnamefont
  {He}}, \bibinfo {author} {\bibfnamefont {S.}~\bibnamefont {Burov}}, \bibinfo
  {author} {\bibfnamefont {R.}~\bibnamefont {Metzler}},\ and\ \bibinfo {author}
  {\bibfnamefont {E.}~\bibnamefont {Barkai}},\ }\bibfield  {title} {\bibinfo
  {title} {{Random Time-Scale Invariant Diffusion and Transport
  Coefficients}},\ }\href {https://doi.org/10.1103/PhysRevLett.101.058101}
  {\bibfield  {journal} {\bibinfo  {journal} {Phys. Rev. Lett.}\ }\textbf
  {\bibinfo {volume} {101}},\ \bibinfo {pages} {058101} (\bibinfo {year}
  {2008})}\BibitemShut {NoStop}%
\bibitem [{\citenamefont {Ikeda}\ \emph {et~al.}(1982)\citenamefont {Ikeda},
  \citenamefont {Kondo},\ and\ \citenamefont {Akimoto}}]{ikeda1982}%
  \BibitemOpen
  \bibfield  {author} {\bibinfo {author} {\bibfnamefont {K.}~\bibnamefont
  {Ikeda}}, \bibinfo {author} {\bibfnamefont {K.}~\bibnamefont {Kondo}},\ and\
  \bibinfo {author} {\bibfnamefont {O.}~\bibnamefont {Akimoto}},\ }\bibfield
  {title} {\bibinfo {title} {{Successive Higher-Harmonic Bifurcations in
  Systems with Delayed Feedback}},\ }\href
  {https://doi.org/10.1103/PhysRevLett.49.1467} {\bibfield  {journal} {\bibinfo
   {journal} {Phys. Rev. Lett.}\ }\textbf {\bibinfo {volume} {49}},\ \bibinfo
  {pages} {1467} (\bibinfo {year} {1982})}\BibitemShut {NoStop}%
\bibitem [{\citenamefont {Chow}\ and\ \citenamefont
  {Mallet-Paret}(1983)}]{chow1983}%
  \BibitemOpen
  \bibfield  {author} {\bibinfo {author} {\bibfnamefont {S.-N.}\ \bibnamefont
  {Chow}}\ and\ \bibinfo {author} {\bibfnamefont {J.}~\bibnamefont
  {Mallet-Paret}},\ }\bibfield  {title} {\bibinfo {title} {{Singularly
  Perturbed Delay-Differential Equations}},\ }\href
  {https://doi.org/10.1016/S0304-0208(08)70968-X} {\bibfield  {journal}
  {\bibinfo  {journal} {North-Holland Mathematics Studies}\ }\textbf {\bibinfo
  {volume} {80}},\ \bibinfo {pages} {7} (\bibinfo {year} {1983})}\BibitemShut
  {NoStop}%
\bibitem [{\citenamefont {Mallet-Paret}\ and\ \citenamefont
  {Nussbaum}(1986)}]{mallet-paret1986}%
  \BibitemOpen
  \bibfield  {author} {\bibinfo {author} {\bibfnamefont {J.}~\bibnamefont
  {Mallet-Paret}}\ and\ \bibinfo {author} {\bibfnamefont {R.~D.}\ \bibnamefont
  {Nussbaum}},\ }\bibfield  {title} {\bibinfo {title} {{Global Continuation and
  Asymptotic Behaviour for Periodic Solutions of a Differential-Delay
  Equation}},\ }\href {https://doi.org/10.1007/BF01790539} {\bibfield
  {journal} {\bibinfo  {journal} {Ann. Mat. Pura Appl.}\ }\textbf {\bibinfo
  {volume} {145}},\ \bibinfo {pages} {33} (\bibinfo {year} {1986})}\BibitemShut
  {NoStop}%
\bibitem [{\citenamefont {Ikeda}\ and\ \citenamefont
  {Matsumoto}(1987)}]{ikeda1987}%
  \BibitemOpen
  \bibfield  {author} {\bibinfo {author} {\bibfnamefont {K.}~\bibnamefont
  {Ikeda}}\ and\ \bibinfo {author} {\bibfnamefont {K.}~\bibnamefont
  {Matsumoto}},\ }\bibfield  {title} {\bibinfo {title} {{High-dimensional
  chaotic behavior in systems with time-delayed feedback}},\ }\href
  {https://doi.org/10.1016/0167-2789(87)90058-3} {\bibfield  {journal}
  {\bibinfo  {journal} {Physica (Amsterdam)}\ }\textbf {\bibinfo {volume}
  {29D}},\ \bibinfo {pages} {223} (\bibinfo {year} {1987})}\BibitemShut
  {NoStop}%
\bibitem [{\citenamefont {Mensour}\ and\ \citenamefont
  {Longtin}(1998)}]{mensour1998}%
  \BibitemOpen
  \bibfield  {author} {\bibinfo {author} {\bibfnamefont {B.}~\bibnamefont
  {Mensour}}\ and\ \bibinfo {author} {\bibfnamefont {A.}~\bibnamefont
  {Longtin}},\ }\bibfield  {title} {\bibinfo {title} {{Chaos control in
  multistable delay-differential equations and their singular limit maps}},\
  }\href {https://doi.org/10.1103/PhysRevE.58.410} {\bibfield  {journal}
  {\bibinfo  {journal} {Phys. Rev. E}\ }\textbf {\bibinfo {volume} {58}},\
  \bibinfo {pages} {410} (\bibinfo {year} {1998})}\BibitemShut {NoStop}%
\bibitem [{\citenamefont {Wolfrum}\ and\ \citenamefont
  {Yanchuk}(2006)}]{wolfrum2006}%
  \BibitemOpen
  \bibfield  {author} {\bibinfo {author} {\bibfnamefont {M.}~\bibnamefont
  {Wolfrum}}\ and\ \bibinfo {author} {\bibfnamefont {S.}~\bibnamefont
  {Yanchuk}},\ }\bibfield  {title} {\bibinfo {title} {{Eckhaus Instability in
  Systems with Large Delay}},\ }\href
  {https://doi.org/10.1103/PhysRevLett.96.220201} {\bibfield  {journal}
  {\bibinfo  {journal} {Phys. Rev. Lett.}\ }\textbf {\bibinfo {volume} {96}},\
  \bibinfo {pages} {220201} (\bibinfo {year} {2006})}\BibitemShut {NoStop}%
\bibitem [{\citenamefont {Adhikari}\ \emph {et~al.}(2008)\citenamefont
  {Adhikari}, \citenamefont {Coutsias},\ and\ \citenamefont
  {McIver}}]{adhikari2008}%
  \BibitemOpen
  \bibfield  {author} {\bibinfo {author} {\bibfnamefont {M.~H.}\ \bibnamefont
  {Adhikari}}, \bibinfo {author} {\bibfnamefont {E.~A.}\ \bibnamefont
  {Coutsias}},\ and\ \bibinfo {author} {\bibfnamefont {J.~K.}\ \bibnamefont
  {McIver}},\ }\bibfield  {title} {\bibinfo {title} {{Periodic solutions of a
  singularly perturbed delay differential equation}},\ }\href
  {https://doi.org/10.1016/j.physd.2008.07.019} {\bibfield  {journal} {\bibinfo
   {journal} {Physica (Amsterdam)}\ }\textbf {\bibinfo {volume} {237D}},\
  \bibinfo {pages} {3307} (\bibinfo {year} {2008})}\BibitemShut {NoStop}%
\bibitem [{\citenamefont {Wolfrum}\ \emph {et~al.}(2010)\citenamefont
  {Wolfrum}, \citenamefont {Yanchuk}, \citenamefont {H\"ovel},\ and\
  \citenamefont {Sch\"oll}}]{wolfrum2010}%
  \BibitemOpen
  \bibfield  {author} {\bibinfo {author} {\bibfnamefont {M.}~\bibnamefont
  {Wolfrum}}, \bibinfo {author} {\bibfnamefont {S.}~\bibnamefont {Yanchuk}},
  \bibinfo {author} {\bibfnamefont {P.}~\bibnamefont {H\"ovel}},\ and\ \bibinfo
  {author} {\bibfnamefont {E.}~\bibnamefont {Sch\"oll}},\ }\bibfield  {title}
  {\bibinfo {title} {{Complex dynamics in delay-differential equations with
  large delay}},\ }\href {https://doi.org/10.1140/epjst/e2010-01343-7}
  {\bibfield  {journal} {\bibinfo  {journal} {Eur. Phys. J. Special Topics}\
  }\textbf {\bibinfo {volume} {191}},\ \bibinfo {pages} {91} (\bibinfo {year}
  {2010})}\BibitemShut {NoStop}%
\bibitem [{\citenamefont {Lichtner}\ \emph {et~al.}(2011)\citenamefont
  {Lichtner}, \citenamefont {Wolfrum},\ and\ \citenamefont
  {Yanchuk}}]{lichtner2011}%
  \BibitemOpen
  \bibfield  {author} {\bibinfo {author} {\bibfnamefont {M.}~\bibnamefont
  {Lichtner}}, \bibinfo {author} {\bibfnamefont {M.}~\bibnamefont {Wolfrum}},\
  and\ \bibinfo {author} {\bibfnamefont {S.}~\bibnamefont {Yanchuk}},\
  }\bibfield  {title} {\bibinfo {title} {{The Spectrum of Delay Differential
  Equations with Large Delay}},\ }\href {https://doi.org/10.1137/090766796}
  {\bibfield  {journal} {\bibinfo  {journal} {SIAM J. Math. Anal.}\ }\textbf
  {\bibinfo {volume} {43}},\ \bibinfo {pages} {788} (\bibinfo {year}
  {2011})}\BibitemShut {NoStop}%
\bibitem [{\citenamefont {Giacomelli}\ \emph {et~al.}(2012)\citenamefont
  {Giacomelli}, \citenamefont {Marino}, \citenamefont {Zaks},\ and\
  \citenamefont {Yanchuk}}]{giacomelli2012}%
  \BibitemOpen
  \bibfield  {author} {\bibinfo {author} {\bibfnamefont {G.}~\bibnamefont
  {Giacomelli}}, \bibinfo {author} {\bibfnamefont {F.}~\bibnamefont {Marino}},
  \bibinfo {author} {\bibfnamefont {M.~A.}\ \bibnamefont {Zaks}},\ and\
  \bibinfo {author} {\bibfnamefont {S.}~\bibnamefont {Yanchuk}},\ }\bibfield
  {title} {\bibinfo {title} {{Coarsening in a bistable system with long-delayed
  feedback}},\ }\href {https://doi.org/10.1209/0295-5075/99/58005} {\bibfield
  {journal} {\bibinfo  {journal} {Europhys. Lett.}\ }\textbf {\bibinfo {volume}
  {99}},\ \bibinfo {pages} {58005} (\bibinfo {year} {2012})}\BibitemShut
  {NoStop}%
\bibitem [{\citenamefont {Marino}\ \emph {et~al.}(2014)\citenamefont {Marino},
  \citenamefont {Giacomelli},\ and\ \citenamefont {Barland}}]{marino2014}%
  \BibitemOpen
  \bibfield  {author} {\bibinfo {author} {\bibfnamefont {F.}~\bibnamefont
  {Marino}}, \bibinfo {author} {\bibfnamefont {G.}~\bibnamefont {Giacomelli}},\
  and\ \bibinfo {author} {\bibfnamefont {S.}~\bibnamefont {Barland}},\
  }\bibfield  {title} {\bibinfo {title} {{Front Pinning and Localized States
  Analogues in Long-Delayed Bistable Systems}},\ }\href
  {https://doi.org/10.1103/PhysRevLett.112.103901} {\bibfield  {journal}
  {\bibinfo  {journal} {Phys. Rev. Lett.}\ }\textbf {\bibinfo {volume} {112}},\
  \bibinfo {pages} {103901} (\bibinfo {year} {2014})}\BibitemShut {NoStop}%
\bibitem [{\citenamefont {Amil}\ \emph {et~al.}(2015)\citenamefont {Amil},
  \citenamefont {Cabeza}, \citenamefont {Masoller},\ and\ \citenamefont
  {Mart\'i}}]{amil2015}%
  \BibitemOpen
  \bibfield  {author} {\bibinfo {author} {\bibfnamefont {P.}~\bibnamefont
  {Amil}}, \bibinfo {author} {\bibfnamefont {C.}~\bibnamefont {Cabeza}},
  \bibinfo {author} {\bibfnamefont {C.}~\bibnamefont {Masoller}},\ and\
  \bibinfo {author} {\bibfnamefont {A.~C.}\ \bibnamefont {Mart\'i}},\
  }\bibfield  {title} {\bibinfo {title} {{Organization and identification of
  solutions in the time-delayed Mackey-Glass model}},\ }\href
  {https://doi.org/10.1063/1.4918593} {\bibfield  {journal} {\bibinfo
  {journal} {Chaos}\ }\textbf {\bibinfo {volume} {25}},\ \bibinfo {pages}
  {043112} (\bibinfo {year} {2015})}\BibitemShut {NoStop}%
\bibitem [{\citenamefont {Lakshmanan}\ and\ \citenamefont
  {Senthilkumar}(2011)}]{lakshmanan2011}%
  \BibitemOpen
  \bibfield  {author} {\bibinfo {author} {\bibfnamefont {M.}~\bibnamefont
  {Lakshmanan}}\ and\ \bibinfo {author} {\bibfnamefont {D.~V.}\ \bibnamefont
  {Senthilkumar}},\ }\href {https://doi.org/10.1007/978-3-642-14938-2} {\emph
  {\bibinfo {title} {{Dynamics of Nonlinear Time-Delay Systems}}}},\ \bibinfo
  {edition} {1st}\ ed.\ (\bibinfo  {publisher} {Springer},\ \bibinfo {address}
  {Berlin Heidelberg},\ \bibinfo {year} {2011})\BibitemShut {NoStop}%
\bibitem [{\citenamefont {Mackey}\ and\ \citenamefont
  {Glass}(1977)}]{mackey1977}%
  \BibitemOpen
  \bibfield  {author} {\bibinfo {author} {\bibfnamefont {M.~C.}\ \bibnamefont
  {Mackey}}\ and\ \bibinfo {author} {\bibfnamefont {L.}~\bibnamefont {Glass}},\
  }\bibfield  {title} {\bibinfo {title} {{Oscillation and Chaos in
  Physiological Control Systems}},\ }\href
  {https://doi.org/10.1126/science.267326} {\bibfield  {journal} {\bibinfo
  {journal} {Science}\ }\textbf {\bibinfo {volume} {197}},\ \bibinfo {pages}
  {287} (\bibinfo {year} {1977})}\BibitemShut {NoStop}%
\bibitem [{\citenamefont {Gushchin}\ and\ \citenamefont
  {K\"uchler}(1999)}]{gushchin1999}%
  \BibitemOpen
  \bibfield  {author} {\bibinfo {author} {\bibfnamefont {A.~A.}\ \bibnamefont
  {Gushchin}}\ and\ \bibinfo {author} {\bibfnamefont {U.}~\bibnamefont
  {K\"uchler}},\ }\bibfield  {title} {\bibinfo {title} {{Asymptotic inference
  for a linear stochastic differential equation with time delay}},\ }\href
  {https://doi.org/10.2307/3318560} {\bibfield  {journal} {\bibinfo  {journal}
  {Bernoulli}\ }\textbf {\bibinfo {volume} {5}},\ \bibinfo {pages} {1059}
  (\bibinfo {year} {1999})}\BibitemShut {NoStop}%
\bibitem [{\citenamefont {Bellman}\ and\ \citenamefont
  {Cooke}(1965)}]{bellman1965}%
  \BibitemOpen
  \bibfield  {author} {\bibinfo {author} {\bibfnamefont {R.}~\bibnamefont
  {Bellman}}\ and\ \bibinfo {author} {\bibfnamefont {K.~L.}\ \bibnamefont
  {Cooke}},\ }\bibfield  {title} {\bibinfo {title} {{On the Computational
  Solution of a Class of Functional Differential Equations}},\ }\href
  {https://doi.org/10.1016/0022-247X(65)90017-X} {\bibfield  {journal}
  {\bibinfo  {journal} {J. Math. Anal. Appl.}\ }\textbf {\bibinfo {volume}
  {12}},\ \bibinfo {pages} {495} (\bibinfo {year} {1965})}\BibitemShut
  {NoStop}%
\bibitem [{\citenamefont {M\"uller}\ \emph {et~al.}(2018)\citenamefont
  {M\"uller}, \citenamefont {Otto},\ and\ \citenamefont
  {Radons}}]{mueller2018}%
  \BibitemOpen
  \bibfield  {author} {\bibinfo {author} {\bibfnamefont {D.}~\bibnamefont
  {M\"uller}}, \bibinfo {author} {\bibfnamefont {A.}~\bibnamefont {Otto}},\
  and\ \bibinfo {author} {\bibfnamefont {G.}~\bibnamefont {Radons}},\
  }\bibfield  {title} {\bibinfo {title} {{Laminar Chaos}},\ }\href
  {https://doi.org/10.1103/PhysRevLett.120.084102} {\bibfield  {journal}
  {\bibinfo  {journal} {Phys. Rev. Lett.}\ }\textbf {\bibinfo {volume} {120}},\
  \bibinfo {pages} {084102} (\bibinfo {year} {2018})}\BibitemShut {NoStop}%
\bibitem [{\citenamefont {M\"uller-Bender}\ \emph {et~al.}(2019)\citenamefont
  {M\"uller-Bender}, \citenamefont {Otto},\ and\ \citenamefont
  {Radons}}]{mueller2019}%
  \BibitemOpen
  \bibfield  {author} {\bibinfo {author} {\bibfnamefont {D.}~\bibnamefont
  {M\"uller-Bender}}, \bibinfo {author} {\bibfnamefont {A.}~\bibnamefont
  {Otto}},\ and\ \bibinfo {author} {\bibfnamefont {G.}~\bibnamefont {Radons}},\
  }\bibfield  {title} {\bibinfo {title} {{Resonant Doppler effect in systems
  with variable delay}},\ }\href {https://doi.org/10.1098/rsta.2018.0119}
  {\bibfield  {journal} {\bibinfo  {journal} {Phil. Trans. R. Soc. A}\ }\textbf
  {\bibinfo {volume} {377}},\ \bibinfo {pages} {20180119} (\bibinfo {year}
  {2019})}\BibitemShut {NoStop}%
\bibitem [{\citenamefont {Bellen}\ and\ \citenamefont
  {Zennaro}(2003)}]{bellen2003}%
  \BibitemOpen
  \bibfield  {author} {\bibinfo {author} {\bibfnamefont {A.}~\bibnamefont
  {Bellen}}\ and\ \bibinfo {author} {\bibfnamefont {M.}~\bibnamefont
  {Zennaro}},\ }\href
  {https://doi.org/10.1093/acprof:oso/9780198506546.001.0001} {\emph {\bibinfo
  {title} {{Numerical Methods for Delay Differential Equations}}}},\ \bibinfo
  {edition} {1st}\ ed.\ (\bibinfo  {publisher} {Oxford University Press},\
  \bibinfo {address} {Oxford},\ \bibinfo {year} {2003})\BibitemShut {NoStop}%
\bibitem [{Note1()}]{Note1}%
  \BibitemOpen
  \bibinfo {note} {We define the solution in one state interval to be laminar
  if the Euclidean distance of the corresponding solution segment from the
  nearest marginally unstable fixed-point solution is smaller than
  $0.05$.}\BibitemShut {Stop}%
\bibitem [{\citenamefont {Karney}(1983)}]{karney1983}%
  \BibitemOpen
  \bibfield  {author} {\bibinfo {author} {\bibfnamefont {C.~F.~F.}\
  \bibnamefont {Karney}},\ }\bibfield  {title} {\bibinfo {title} {{Long-time
  correlations in the stochastic regime}},\ }\href
  {https://doi.org/10.1016/0167-2789(83)90232-4} {\bibfield  {journal}
  {\bibinfo  {journal} {Physica (Amsterdam)}\ }\textbf {\bibinfo {volume}
  {8D}},\ \bibinfo {pages} {360} (\bibinfo {year} {1983})}\BibitemShut
  {NoStop}%
\bibitem [{\citenamefont {Ichikawa}\ \emph {et~al.}(1987)\citenamefont
  {Ichikawa}, \citenamefont {Kamimura},\ and\ \citenamefont
  {Hatori}}]{ichikawa1987}%
  \BibitemOpen
  \bibfield  {author} {\bibinfo {author} {\bibfnamefont {Y.~H.}\ \bibnamefont
  {Ichikawa}}, \bibinfo {author} {\bibfnamefont {T.}~\bibnamefont {Kamimura}},\
  and\ \bibinfo {author} {\bibfnamefont {T.}~\bibnamefont {Hatori}},\
  }\bibfield  {title} {\bibinfo {title} {{Stochastic diffusion in the standard
  map}},\ }\href {https://doi.org/10.1016/0167-2789(87)90060-1} {\bibfield
  {journal} {\bibinfo  {journal} {Physica (Amsterdam)}\ }\textbf {\bibinfo
  {volume} {29D}},\ \bibinfo {pages} {247} (\bibinfo {year}
  {1987})}\BibitemShut {NoStop}%
\bibitem [{\citenamefont {Zaslavsky}(2002)}]{zaslavsky2002}%
  \BibitemOpen
  \bibfield  {author} {\bibinfo {author} {\bibfnamefont {G.~M.}\ \bibnamefont
  {Zaslavsky}},\ }\bibfield  {title} {\bibinfo {title} {{Dynamical traps}},\
  }\href {https://doi.org/10.1016/S0167-2789(02)00516-X} {\bibfield  {journal}
  {\bibinfo  {journal} {Physica D}\ }\textbf {\bibinfo {volume} {168--169}},\
  \bibinfo {pages} {292} (\bibinfo {year} {2002})}\BibitemShut {NoStop}%
\bibitem [{\citenamefont {Farmer}(1982)}]{farmer1982}%
  \BibitemOpen
  \bibfield  {author} {\bibinfo {author} {\bibfnamefont {J.~D.}\ \bibnamefont
  {Farmer}},\ }\bibfield  {title} {\bibinfo {title} {{Chaotic attractors of an
  infinite-dimensional dynamical system}},\ }\href
  {https://doi.org/10.1016/0167-2789(82)90042-2} {\bibfield  {journal}
  {\bibinfo  {journal} {Physica (Amsterdam)}\ }\textbf {\bibinfo {volume}
  {4D}},\ \bibinfo {pages} {366} (\bibinfo {year} {1982})}\BibitemShut
  {NoStop}%
\bibitem [{\citenamefont {Amann}\ \emph {et~al.}(2007)\citenamefont {Amann},
  \citenamefont {Sch\"oll},\ and\ \citenamefont {Just}}]{amann2007}%
  \BibitemOpen
  \bibfield  {author} {\bibinfo {author} {\bibfnamefont {A.}~\bibnamefont
  {Amann}}, \bibinfo {author} {\bibfnamefont {E.}~\bibnamefont {Sch\"oll}},\
  and\ \bibinfo {author} {\bibfnamefont {W.}~\bibnamefont {Just}},\ }\bibfield
  {title} {\bibinfo {title} {{Some basic remarks on eigenmode expansions of
  time-delay dynamics}},\ }\href {https://doi.org/10.1016/j.physa.2005.12.073}
  {\bibfield  {journal} {\bibinfo  {journal} {Physica A}\ }\textbf {\bibinfo
  {volume} {373}},\ \bibinfo {pages} {191} (\bibinfo {year}
  {2007})}\BibitemShut {NoStop}%
\bibitem [{\citenamefont {Redmond}\ \emph {et~al.}(2002)\citenamefont
  {Redmond}, \citenamefont {LeBlanc},\ and\ \citenamefont
  {Longtin}}]{redmond2002}%
  \BibitemOpen
  \bibfield  {author} {\bibinfo {author} {\bibfnamefont {B.~F.}\ \bibnamefont
  {Redmond}}, \bibinfo {author} {\bibfnamefont {V.~G.}\ \bibnamefont
  {LeBlanc}},\ and\ \bibinfo {author} {\bibfnamefont {A.}~\bibnamefont
  {Longtin}},\ }\bibfield  {title} {\bibinfo {title} {{Bifurcation analysis of
  a class of first-order nonlinear delay-differential equations with
  reflectional symmetry}},\ }\href
  {https://doi.org/10.1016/S0167-2789(02)00423-2} {\bibfield  {journal}
  {\bibinfo  {journal} {Physica D}\ }\textbf {\bibinfo {volume} {166}},\
  \bibinfo {pages} {131} (\bibinfo {year} {2002})}\BibitemShut {NoStop}%
\bibitem [{Note2()}]{Note2}%
  \BibitemOpen
  \bibinfo {note} {The solution in one state interval is identified as a
  plateau if the median of all derivatives of the solution inside the state
  interval is smaller than 0.01. A plateau belongs to a doubly-laminar phase if
  the value of the plateau, i.e., the value of the solution taken in the middle
  of a state interval is closer than 0.05 to the nearest marginally unstable
  fixed point. Otherwise, it belongs to a chaotic-laminar phase.}\BibitemShut
  {Stop}%
\bibitem [{\citenamefont {Hart}\ \emph {et~al.}(2019)\citenamefont {Hart},
  \citenamefont {Roy}, \citenamefont {M\"uller-Bender}, \citenamefont {Otto},\
  and\ \citenamefont {Radons}}]{hart2019}%
  \BibitemOpen
  \bibfield  {author} {\bibinfo {author} {\bibfnamefont {J.~D.}\ \bibnamefont
  {Hart}}, \bibinfo {author} {\bibfnamefont {R.}~\bibnamefont {Roy}}, \bibinfo
  {author} {\bibfnamefont {D.}~\bibnamefont {M\"uller-Bender}}, \bibinfo
  {author} {\bibfnamefont {A.}~\bibnamefont {Otto}},\ and\ \bibinfo {author}
  {\bibfnamefont {G.}~\bibnamefont {Radons}},\ }\bibfield  {title} {\bibinfo
  {title} {{Laminar Chaos in Experiments: Nonlinear Systems with Time-Varying
  Delays and Noise}},\ }\href {https://doi.org/10.1103/PhysRevLett.123.154101}
  {\bibfield  {journal} {\bibinfo  {journal} {Phys. Rev. Lett.}\ }\textbf
  {\bibinfo {volume} {123}},\ \bibinfo {pages} {154101} (\bibinfo {year}
  {2019})}\BibitemShut {NoStop}%
\bibitem [{\citenamefont {M\"uller-Bender}\ \emph {et~al.}(2020)\citenamefont
  {M\"uller-Bender}, \citenamefont {Otto}, \citenamefont {Radons},
  \citenamefont {Hart},\ and\ \citenamefont {Roy}}]{mueller2020}%
  \BibitemOpen
  \bibfield  {author} {\bibinfo {author} {\bibfnamefont {D.}~\bibnamefont
  {M\"uller-Bender}}, \bibinfo {author} {\bibfnamefont {A.}~\bibnamefont
  {Otto}}, \bibinfo {author} {\bibfnamefont {G.}~\bibnamefont {Radons}},
  \bibinfo {author} {\bibfnamefont {J.~D.}\ \bibnamefont {Hart}},\ and\
  \bibinfo {author} {\bibfnamefont {R.}~\bibnamefont {Roy}},\ }\bibfield
  {title} {\bibinfo {title} {{Laminar chaos in experiments and nonlinear
  delayed Langevin equations: A time series analysis toolbox for the detection
  of laminar chaos}},\ }\href {https://doi.org/10.1103/PhysRevE.101.032213}
  {\bibfield  {journal} {\bibinfo  {journal} {Phys. Rev. E}\ }\textbf {\bibinfo
  {volume} {101}},\ \bibinfo {pages} {032213} (\bibinfo {year}
  {2020})}\BibitemShut {NoStop}%
\bibitem [{\citenamefont {Otto}\ \emph {et~al.}(2017)\citenamefont {Otto},
  \citenamefont {M\"uller},\ and\ \citenamefont {Radons}}]{otto2017}%
  \BibitemOpen
  \bibfield  {author} {\bibinfo {author} {\bibfnamefont {A.}~\bibnamefont
  {Otto}}, \bibinfo {author} {\bibfnamefont {D.}~\bibnamefont {M\"uller}},\
  and\ \bibinfo {author} {\bibfnamefont {G.}~\bibnamefont {Radons}},\
  }\bibfield  {title} {\bibinfo {title} {{Universal Dichotomy for Dynamical
  Systems with Variable Delay}},\ }\href
  {https://doi.org/10.1103/PhysRevLett.118.044104} {\bibfield  {journal}
  {\bibinfo  {journal} {Phys. Rev. Lett.}\ }\textbf {\bibinfo {volume} {118}},\
  \bibinfo {pages} {044104} (\bibinfo {year} {2017})}\BibitemShut {NoStop}%
\bibitem [{\citenamefont {M\"uller}\ \emph {et~al.}(2017)\citenamefont
  {M\"uller}, \citenamefont {Otto},\ and\ \citenamefont
  {Radons}}]{mueller2017}%
  \BibitemOpen
  \bibfield  {author} {\bibinfo {author} {\bibfnamefont {D.}~\bibnamefont
  {M\"uller}}, \bibinfo {author} {\bibfnamefont {A.}~\bibnamefont {Otto}},\
  and\ \bibinfo {author} {\bibfnamefont {G.}~\bibnamefont {Radons}},\
  }\bibfield  {title} {\bibinfo {title} {{From dynamical systems with
  time-varying delay to circle maps and Koopman operators}},\ }\href
  {https://doi.org/10.1103/PhysRevE.95.062214} {\bibfield  {journal} {\bibinfo
  {journal} {Phys. Rev. E}\ }\textbf {\bibinfo {volume} {95}},\ \bibinfo
  {pages} {062214} (\bibinfo {year} {2017})}\BibitemShut {NoStop}%
\bibitem [{\citenamefont {Manneville}\ and\ \citenamefont
  {Pomeau}(1979)}]{manneville1979}%
  \BibitemOpen
  \bibfield  {author} {\bibinfo {author} {\bibfnamefont {P.}~\bibnamefont
  {Manneville}}\ and\ \bibinfo {author} {\bibfnamefont {Y.}~\bibnamefont
  {Pomeau}},\ }\bibfield  {title} {\bibinfo {title} {{Intermittency and the
  Lorenz model}},\ }\href {https://doi.org/10.1016/0375-9601(79)90255-X}
  {\bibfield  {journal} {\bibinfo  {journal} {Phys. Lett. A}\ }\textbf
  {\bibinfo {volume} {75}},\ \bibinfo {pages} {1} (\bibinfo {year}
  {1979})}\BibitemShut {NoStop}%
\bibitem [{\citenamefont {Pomeau}\ and\ \citenamefont
  {Manneville}(1980)}]{pomeau1980}%
  \BibitemOpen
  \bibfield  {author} {\bibinfo {author} {\bibfnamefont {Y.}~\bibnamefont
  {Pomeau}}\ and\ \bibinfo {author} {\bibfnamefont {P.}~\bibnamefont
  {Manneville}},\ }\bibfield  {title} {\bibinfo {title} {{Intermittent
  Transition to Turbulence in Dissipative Dynamical Systems}},\ }\href
  {https://doi.org/10.1007/BF01197757} {\bibfield  {journal} {\bibinfo
  {journal} {Commun. Math. Phys.}\ }\textbf {\bibinfo {volume} {74}},\ \bibinfo
  {pages} {189} (\bibinfo {year} {1980})}\BibitemShut {NoStop}%
\bibitem [{\citenamefont {Manneville}(1980)}]{manneville1980}%
  \BibitemOpen
  \bibfield  {author} {\bibinfo {author} {\bibfnamefont {P.}~\bibnamefont
  {Manneville}},\ }\bibfield  {title} {\bibinfo {title} {{Intermittency,
  self-similarity and 1/f spectrum in dissipative dynamical systems}},\ }\href
  {https://doi.org/10.1051/jphys:0198000410110123500} {\bibfield  {journal}
  {\bibinfo  {journal} {J. Phys. France}\ }\textbf {\bibinfo {volume} {41}},\
  \bibinfo {pages} {1235} (\bibinfo {year} {1980})}\BibitemShut {NoStop}%
\bibitem [{Note3()}]{Note3}%
  \BibitemOpen
  \bibinfo {note} {The Kaplan-Yorke dimension can be directly calculated from
  the Lyapunov spectrum, i.e., the set of all Lyapunov exponents. The latter
  were defined as time averages of the expansion rates inside only the
  chaotic-laminar and turbulent phases. They were numerically calculated by
  using the method in \cite {farmer1982} and in order to supress fluctuations,
  the obtained exponents were averaged over 1024 initial
  functions.}\BibitemShut {Stop}%
\bibitem [{\citenamefont {Zumofen}\ and\ \citenamefont
  {Klafter}(1995)}]{zumofen1995}%
  \BibitemOpen
  \bibfield  {author} {\bibinfo {author} {\bibfnamefont {G.}~\bibnamefont
  {Zumofen}}\ and\ \bibinfo {author} {\bibfnamefont {J.}~\bibnamefont
  {Klafter}},\ }\bibfield  {title} {\bibinfo {title} {{Laminar--localized-phase
  coexistence in dynamical systems}},\ }\href
  {https://doi.org/10.1103/PhysRevE.51.1818} {\bibfield  {journal} {\bibinfo
  {journal} {Phys. Rev. E}\ }\textbf {\bibinfo {volume} {51}},\ \bibinfo
  {pages} {1818} (\bibinfo {year} {1995})}\BibitemShut {NoStop}%
\bibitem [{\citenamefont {Liu}\ \emph {et~al.}(2022)\citenamefont {Liu},
  \citenamefont {Jin}, \citenamefont {Bao},\ and\ \citenamefont
  {Chen}}]{liu2022}%
  \BibitemOpen
  \bibfield  {author} {\bibinfo {author} {\bibfnamefont {J.}~\bibnamefont
  {Liu}}, \bibinfo {author} {\bibfnamefont {Y.}~\bibnamefont {Jin}}, \bibinfo
  {author} {\bibfnamefont {J.-D.}\ \bibnamefont {Bao}},\ and\ \bibinfo {author}
  {\bibfnamefont {X.}~\bibnamefont {Chen}},\ }\bibfield  {title} {\bibinfo
  {title} {{Coexistence of ergodicity and nonergodicity in the aging two-state
  random walks}},\ }\href {https://doi.org/10.1039/d2sm01093c} {\bibfield
  {journal} {\bibinfo  {journal} {Soft Matter}\ }\textbf {\bibinfo {volume}
  {18}},\ \bibinfo {pages} {8687} (\bibinfo {year} {2022})}\BibitemShut
  {NoStop}%
\bibitem [{\citenamefont {Einstein}(1905)}]{einstein1905}%
  \BibitemOpen
  \bibfield  {author} {\bibinfo {author} {\bibfnamefont {A.}~\bibnamefont
  {Einstein}},\ }\bibfield  {title} {\bibinfo {title} {{\"Uber die von der
  molekularkinetischen Theorie der W\"arme geforderte Bewegung von in ruhenden
  Fl\"ussigkeiten suspendierten Teilchen}},\ }\href
  {https://doi.org/10.1002/andp.19053220806} {\bibfield  {journal} {\bibinfo
  {journal} {Ann. Phys. (Berlin)}\ }\textbf {\bibinfo {volume} {322}},\
  \bibinfo {pages} {549} (\bibinfo {year} {1905})}\BibitemShut {NoStop}%
\bibitem [{\citenamefont {Geisel}\ \emph {et~al.}(1985)\citenamefont {Geisel},
  \citenamefont {Nierwetberg},\ and\ \citenamefont {Zacherl}}]{geisel1985}%
  \BibitemOpen
  \bibfield  {author} {\bibinfo {author} {\bibfnamefont {T.}~\bibnamefont
  {Geisel}}, \bibinfo {author} {\bibfnamefont {J.}~\bibnamefont
  {Nierwetberg}},\ and\ \bibinfo {author} {\bibfnamefont {A.}~\bibnamefont
  {Zacherl}},\ }\bibfield  {title} {\bibinfo {title} {{Accelerated Diffusion in
  Josephson Junctions and Related Chaotic Systems}},\ }\href
  {https://doi.org/10.1103/PhysRevLett.54.616} {\bibfield  {journal} {\bibinfo
  {journal} {Phys. Rev. Lett.}\ }\textbf {\bibinfo {volume} {54}},\ \bibinfo
  {pages} {616} (\bibinfo {year} {1985})}\BibitemShut {NoStop}%
\end{thebibliography}%

\end{document}